\documentstyle[12pt]{article}

\def\hybrid{\topmargin 0pt      \oddsidemargin 0pt
        \headheight 0pt \headsep 0pt

       \textwidth 6.5in        
       \textheight 9in         
        \marginparwidth 0.0in
        \parskip 5pt plus 1pt   \jot = 1.5ex}
\catcode`\@=11
\def\marginnote#1{}

\newcount\hour
\newcount\minute
\newtoks\amorpm
\hour=\time\divide\hour by60
\minute=\time{\multiply\hour by60 \global\advance\minute by-\hour}
\edef\standardtime{{\ifnum\hour<12 \global\amorpm={am}%
        \else\global\amorpm={pm}\advance\hour by-12 \fi
        \ifnum\hour=0 \hour=12 \fi
        \number\hour:\ifnum\minute<10 0\fi\number\minute\the\amorpm}}
\edef\militarytime{\number\hour:\ifnum\minute<10 0\fi\number\minute}

\def\draftlabel#1{{\@bsphack\if@filesw {\let\thepage\relax
   \xdef\@gtempa{\write\@auxout{\string
      \newlabel{#1}{{\@currentlabel}{\thepage}}}}}\@gtempa
   \if@nobreak \ifvmode\nobreak\fi\fi\fi\@esphack}
        \gdef\@eqnlabel{#1}}
\def\@eqnlabel{}
\def\@vacuum{}

\def\draftmarginnote#1{\marginpar{\raggedright\scriptsize\tt#1}}

\def\draftlabel#1{{\@bsphack\if@filesw {\let\thepage\relax
   \xdef\@gtempa{\write\@auxout{\string
      \newlabel{#1}{{\@currentlabel}{\thepage}}}}}\@gtempa
   \if@nobreak \ifvmode\nobreak\fi\fi\fi\@esphack}
        \gdef\@eqnlabel{#1}}
\def\@eqnlabel{}
\def\@vacuum{}
\def\draftmarginnote#1{\marginpar{\raggedright\scriptsize\tt#1}}

\def\draft{\oddsidemargin -.5truein
        \def\@oddfoot{\sl preliminary draft \hfil
        \rm\thepage\hfil\sl\today\quad\militarytime}
        \let\@evenfoot\@oddfoot \overfullrule 3pt
        \let\label=\draftlabel
        \let\marginnote=\draftmarginnote
   \def\@eqnnum{(\theequation)\rlap{\kern\marginparsep\tt\@eqnlabel}%
\global\let\@eqnlabel\@vacuum}  }

\def\numberbysection{\@addtoreset{equation}{section}
        \def\theequation{\thesection.\arabic{equation}}}

\def\underline#1{\relax\ifmmode\@@underline#1\else
        $\@@underline{\hbox{#1}}$\relax\fi}

\def\titlepage{\@restonecolfalse\if@twocolumn\@restonecoltrue\onecolumn
     \else \newpage \fi \thispagestyle{empty}\c@page\z@
        \def\thefootnote{\fnsymbol{footnote}} }

\def\endtitlepage{\if@restonecol\twocolumn \else  \fi
        \def\thefootnote{\arabic{footnote}}
        \setcounter{footnote}{0}}  
\relax

\numberbysection
\hybrid

\def\beq{\begin{equation}}
\def\eeq{\end{equation}}
\def\p{\partial}
\def\G{\Gamma}

\def\s{\sigma}

\def\a{\alpha}

\def\bea{\begin{eqnarray}}
\def\eea{\end{eqnarray}}

\begin{document}

\input epsf

\begin{titlepage}

\title{Periodic and almost periodic potentials in the inverse problems}

\author{I.Krichever \thanks{Columbia University, 2990 Broadway,
New York, NY 10027, USA and Landau Institute
for Theoretical Physics, Kosygina str. 2, 117940 Moscow, Russia; e-mail:
krichev@math.columbia.edu. Research supported in part by National Science
Foundation under the grant DMS-98-02577}
\and
S.P.Novikov\thanks{University of Maryland, Colledge Park and Landau Institute
for Theoretical Physics, Kosygina str. 2, 117940 Moscow, Russia; e-mail:
novikov@ipst.umd.edu}}
\date{February, 1999}
\maketitle


\end{titlepage}
\newpage

\section{Introduction}

To begin with we are going to consider the inverse spectral problem for
one-dimensional Schr\"odinger operator with periodic potential. In the late
60-s the famous discovery of the Inverse Scattering Transform for the KdV
equation was done. A periodic analog of this transform
was found in 1974. It is based on the solution of the following inverse
spectral problem:

\noindent {\it to describe effectively  the ''isospectral manifold''
of all the potentials with a given spectrum on the line (i.e.  the
spectrum of Schr\"odinger operator acting in the Hilbert space of square
integrable complex-valued functions on the line $R$)}.

As everybody working in the quantum solid state physics knows,
this spectrum generically is a union of infinite number of intervals
(allowed bands) on the energy line $\epsilon$. The complementary part on the
energy line is also a union of infinite number of intervals (gaps or
forbidden bands) whose lengths tend to zero for $\epsilon\to+\infty$.

The periodic problem was solved in 1974-75 for the so-called {\it
finite-gap potentials} that have only a finite number of gaps. Any periodic
potential can be approximated by the finite-gap ones. This solution involves
combination of the theory of Riemann surfaces and their $\theta$-functions,
Hamiltonian dynamics of special completely integrable systems and
the spectral theory of Schr\"odinger operator.  The mathematical technique
used was (and still is) unusual for the community of physicists. Later
the necessity to use this kind of mathematics appeared also in other branches
of mathematical and theoretical physics (for example, in the  string theory,
matrix models, and supersymmetric Yang-Mills theory
\cite{DW,DS,dub1,dub2,gkmmm,gn,kr_tau,kph1,kph2,mart,martW,marsh,mmm,sw1,sw2}).
It seems to the authors this technique in future will be needed to the broad
community of theoretical physicists.

Integrability of the famous KdV equation $u_t=6uu_x+u_{xxx}$ was discovered
in 1965-68 (see \cite{KZ,GGKM,L}) for the rapidly decreasing initial data on
the line $x$. Exact solutions for the KdV equation expressing $u(x,t)$ through
the {\it inverse scattering data} of the Schr\"odinger operator
$L=-\partial^2_x+u(x,0)$ were found. This procedure has been called the Inverse
Scattering Transform (IST).  It was extended later for some other highly
nontrivial $(1+1)$-systems including such famous systems as Nonlinear
Schr\"odinger
$NS_{\pm}:i\psi_t=-\psi_{xx}\pm |\psi |^2\psi$ and Sine(Sinh)-Gordon equations
$SG:u_{xt}=\sin u$ or $u_{xt}=\sinh u$. Note, that for the SG equation a large
family of exact solutions was already constructed in  XIX
century by Bianchi, Lie and B\"acklund (see in \cite{ZS,Lm,AKNS}).
Beginning with 1974,
several $(2+1)$-dimensional physically interesting systems have been
discovered as integrable by the IST procedure. The most famous of them is
the KP system (see \cite{ZS1,Dr}).

It is necessary to emphasize that the IST procedure in its original form
can not be applied to the solution of the periodic problem (i.e. $u(x,t)$ is
periodic in the variable $x$). This problem was solved on the
base of the new approach proposed in \cite{N}, in the works
\cite{DN,DN1,D,IM,IM1,L2,MvM}(see the surveys \cite{DMN,K1,KN}).
An extension of this method to $(2+1)$-systems was found in \cite{K,Kf,K1}.
New development of this approach associated with two-dimensional
Schr\"odinger operator was started in 1976
(see \cite{DKN,KN,VN,NV,Ksp,Ksit,Kort}).

Complete detailed description of the solution  of
the periodic problem can be
found in the surveys, Encyclopedia articles \cite{DKN1}, and in the book
\cite{NMPZ}. We are going to present here basic ideas of this theory in the
simplest form possible.  Let us point out that KdV as well as other nontrivial
{\it completely integrable by IST} PDE systems  are indeed completely
integrable in any reasonable sense for rapidly decreasing or periodic
(quasi-periodic) boundary conditions, only.
In fact, even that is well established for few of them.
For example, for the KdV any periodic solution can be approximated
by the finite-gap solutions. This statement easily follows from the
theory of finite-gap potentials if we do not try to preserve the period, i.e.
in the class of all quasi-periodic finite-gap potentials. The approximation
of any periodic potential by the finite-gap potentials with exactly  the
same period, was constructed on the base of other approach developed in
\cite{MO}. The extension of the theory of Riemann surfaces and
$\theta$-functions
to the specific class of  surfaces of the infinite genus associated with
periodic Schr\"odinger operator, was done in \cite{McKT}. This theory
is a beautiful description of the infinite limit. However,
it seems that all fundamental properties of $\theta$-functions
associated with the complex continuation of variables, are lost in this limit.
It is interesting to point out that analogous (but more complicated)
theory of Riemann surfaces of the infinite genus was developed later in
\cite{Ksp} for the periodic 2D Schr\"odinger operators.

Outside these functional classes almost no effective
information is known. Beautiful methods have been developed also for the
studies of
the special self-similar and ''string-type'' solutions, but in most cases they
lead to the very hard analytical problems associated with the famous Painleve
equations and their generalizations (\cite{MTW,FN,Jap,Jap1,IN,KKop,GN,DZ}).

\section {Rapidly decreasing potentials and GGKM Procedure. B\"acklund
Transformations}

Let us recall the basic information about the IST method for KdV.  We start from
the so-called {\it Lax Representation} for KdV (see \cite{L}). The Heisenberg
type equation for the Schr\"odinger operator $L$
\beq
L_t=[L,A]=LA-AL,           \label{LR}
\eeq
\beq
L=-\partial_x^2+u, \ \ A=-4\partial_x^3+3(u\partial_x +\partial_x u),
\eeq
is equivalent to the identity (KdV equation)
\beq
u_t=6uu_x+u_{xxx}.                   \label{KdV}
\eeq
For this reason, any KdV type equations admitting some analog of the Lax
representation is called {\it isospectral deformations}. The existence of such
deformations indicates the possibility of effective solution of the Inverse
Scattering Problem for the operator $L$.

For the rapidly enough decreasing functions $u(x,t)\to 0, x\to \pm\infty$
we define two bases of solutions ($t$ is fixed):
\begin{eqnarray}                                \label{As}
\phi_{\pm}(x,t;k)\sim \exp^{\pm ikx},\ & x\to -\infty, &
L\phi_{\pm}=\lambda\phi_{\pm},\\
\psi_{\pm}(x,t;k)\sim \exp^{\pm ikx}, \ & x\to +\infty, &
L\psi_{\pm}=\lambda\psi_{\pm}, \ k^2=\lambda. \end{eqnarray}
By definition, monodromy matrix $T$ connects these two bases,
$T\phi =\psi$ for the column vectors $\phi=(\phi_+,\phi_-), \
\psi=(\psi_+,\psi_-)$:
\begin{eqnarray}                                          \label{Mon}
T=\left(\begin{array}{cc}
a&b\\
c&d
\end{array}\right),\ \psi_+=a\phi_++b\phi_-,\ \psi_-=c\phi_++d\phi_-\ .
\end{eqnarray}
A conservation of the Wronskian implies that $\det T=ad-bc=1$. For
the real values of $k$ or $\lambda > 0$ we have $a=\bar{d}, c=\bar{b}$.
Therefore, $|a|^2-|b|^2=1$. The whole set of the so-called inverse scattering
data can be extracted from the monodromy matrix $T$ if it is well-defined for
all complex values of $k$.  The so-called {\it scattering matrix}
is constructed from $T$ for the real $k$. Its entries are the
{\it transmission} coefficient $1/a$ and
the {\it reflection} coefficient $\bar{b}/a$.
The property $b=0$ for all real values
of $k$ characterizes reflexionless or multisoliton potentials. For all
rapidly decreasing potentials the matrix element $a(k)$ is well-defined for
complex $k$ such that ${\rm Im}\  k>0$ and $a\to 1$ for $k\to \infty, \ {\rm
Im}\  k>0$.  There is only a finite number of purely imaginary zeroes
$a(k_n)=0$ in this domain. They correspond to the discrete spectrum
$\lambda_n=k_n^2<0$.

\noindent
The famous result of \cite{GGKM} (GGKM procedure) easily follows from the Lax
representation, which implies the equation:
\begin{eqnarray}                           \label{TL}
T_t=[T,\Lambda] ,\ \  \Lambda=\left (\begin{array}{cc} -4ik&0\\
0&4ik \end{array}\right )
\end{eqnarray}

\noindent This result was formulated as a set of the following GGKM formulas
\begin{eqnarray}
a_t=0,\ b_t=-8(ik)^3=-c_t, \ d_t=0.                  \label{GGKM}
\end{eqnarray}
The latter equations give a full description of KdV dynamics in these variables
because any rapidly decreasing potential can be reconstructed from the
inverse scattering data. A special family of the reflexionless potentials where $b=0$
for real values of $k$, leads to the so-called multisoliton solutions for KdV
equation (see \cite{NMPZ}.)

The multisoliton solutions can be also directly obtained with the help of the
elementary substitutions (B\"aclund Transformations) transforming any
solution of the KdV into another solution:  let $u$ be a solution of the
KdV equation and $v$ be a solution of the Ricatti equation $\alpha+u=v_x+v^2$
with the initial value independent of time. The new function
$\tilde{u}=-v_x+v^2$ satisfies the KdV equation.
Starting from the trivial solution $u=u_0=0$ we construct a sequence of
potentials $\tilde{u}_{n-1}=u_n,n> 0, $ given by the B\"acklund Transformation.
We choose parameters $\alpha_n, \ \alpha_1 >\alpha_2> \ldots
>\alpha_n >\ldots$, and take the real nonzero functions
$f_n\to \infty, \ x\to \pm \infty,\ -f_{nxx}+u_{n-1}f_n=\alpha_nf_n$, which
define $v_n=(\log f_n)_x$.
Every such sequence leads to the
multisoliton reflexionless potential:
\beq
u_0=0, \ u_1=-\frac{2\alpha}{ch^2(\sqrt{\alpha}(x-x_0)+\beta t)},\ \ldots
\label{sol}
\eeq
In terms of the Schr\"odinger operator this transformation (invented by L.Euler
in 1742) is called  Darboux Transformation. The operator $L$ can be factorized
$$L=-\partial^2+u=-(\partial+v)(\partial-v).$$
Using  the non-commutativity of these
factors, we define the Darboux Transformation for the operator and its
eigenfunction in the following way:
\beq\label{Dar}
\tilde{L}=-\partial^2+\tilde{u}=-(\partial-v)(\partial+v),\
\tilde{\psi}=(\partial+v)\psi\ .
\eeq
These transformations can be considered as some kind of {\it discrete spectral
symmetries} for Schr\"odinger operators.  They preserve a spectrum of the
operator $L$ (maybe except for one eigenfunction).

\section{KdV Hierarchy. Integrals of motion. Hamiltonian formalism.}

The local integrals for the KdV equation can be constructed with the help of the
Schr\"odinger operator. Consider the associated Riccati equation $v_x+v^2=u-k^2$
and find the solution for it as a formal series in the variable $k$:
\beq
v(x,k)=ik+\sum_{n=1}^{\infty}v_n(x)(ik)^{-n},   \label{As1}
\eeq
where all $v_n$ are polynomials in the variables $u,\ u_x, \ldots$
The integral along  the line $x$ is a $k$-dependent constant of motion for the
KdV equation
\beq \p_t \left(\int v(x,k)dx\right)=0 ,\ u_t=6uu_x+u_{xxx}. \label{Int}
\eeq
For the real potentials $u(x,t)$ and real $k$ we can see that the
imaginary part of $v(x,k)-ik$ is a total derivative. The remaining quantities
in the expansion :
\beq
\int (v(x,k)-ik)dx=\sum_{n\geq 1} \int v_n(x)(ik)^{-n}dx. \label{Iloc},
\eeq
define local integrals of motion
\beq
I_n=c_n\int v_{2n+3}(x)dx,\ n=-1,0,1,2,\ldots,
\eeq
where $c_n$ are constants.
After the proper choice of the constants $c_n$ we have
\beq
I_{-1}=\int udx, I_0=\int u^2 dx, I_1=\int (u^2_x/2+u^3)dx,\ldots,
\eeq
\beq
I_{nt}=0, \ \ u_t=6uu_x+u_{xxx}.
\eeq
Let us introduce a GZF Poisson bracket \cite{G,ZF} on the space of functions
\beq
\{ u(x),u(y)\}=\delta' (x-y)). \label{GZF}
\eeq
Then any functional $H$ (Hamiltonian) defines the corresponding
Hamiltonian system
\beq
u_t=\partial_x\left(\frac{\delta H}{\delta u(x)}\right).     \label{Gform}
\eeq
For the case $H=I_{-1}$ we get a trivial flow (i.e. this integral is a Casimir
for the GZF bracket).  For $H=I_0$ we are coming to the $x$-translations
$u_t=u_x$. Let us call this equation $KdV_0$. For the case $H=I_1$
we have the ordinary $KdV=KdV_1$. Higher integrals give us the
equations $KdV_n$ of the order $2n+1$ admitting the Lax representations with the
same Schr\"odinger operator $L$ but with the differential operators
$A_n=(const)\partial_x^{2n+1}+\dots$ :
\beq
u_{t_n}=\partial_x \left(\frac{\delta I_n}{\delta u(x)}\right)=[L,A_n].
\label{LR2}
\eeq
\noindent In particular, $A_0=\partial_x,\ A_1=A.$  Nice
formula for all operators $A_n$ can be extracted from \cite{GD}.

Let $L=-{\cal L}^2$, where
${\cal L}=\partial_x+\sum_{k\geq 1}a_k(u,u_x,\ldots )\partial_x^{-k}$. Here all
$a_k$ are polynomials in the variables $u,u_x,\ldots $ and
$\partial_x^{-1}a=\sum_{n\geq 0}(-1)^na^{(n)}\partial_x^{-n-1}$ for the
composition of the operator $\partial^{-1}$ and multiplication operator by
$a$. By definition,
\beq A_n=({\cal L}^{2n+1})_+=(L^{2n+1/2})_+,
\eeq

\noindent where the sign $+$ means omitting of all strictly negative powers of
$\partial_x $.

All higher $KdV_n$ systems can be integrated by the same IST procedure for the
class of rapidly decreasing functions.  In particular, GGKM equations for the
scattering data (or monodromy matrix) have the form
\beq
T_{t_n}=[T,\Lambda_n ],\ \Lambda_n=(const)\left ( \begin{array}{cc}
(ik)^{2n+1}&0\\0&(-ik)^{2n+1}\end{array}\right ).
\label{GGKM1}
\eeq
The latter result also implies that all these flows commute with each other.
Hence, we get the following conclusion without any calculation:

{\it integrals $I_n$ have zero Poisson brackets},
\beq
\{I_n,I_m\}=\int\frac{\delta I_n}{\delta u(x)}\partial_x \frac{\delta
I_m}{\delta u(x)} dx=0.
\label{PB}
\eeq
A generalization of the GZF Poisson bracket for the
isospectral deformations of the higher order (scalar) Lax operators
$L$ was found in \cite{GD2}.

It should be emphasized that there exists a family of local field-theoretical
Poisson brackets (LM-brackets, \cite{Ma}) describing the KdV theory:
\beq
B= B_{(\lambda,\mu)}= \lambda\partial_x +\mu (-\partial_x^3+4u\partial_x+2u_x)=
\lambda B_0+\mu B_1,
\label{PB2}
\eeq
\beq
\{u(x),u(y)\}_{\lambda,\mu}=B\delta(x-y),
\eeq
\beq
\{I_n,I_m\}_{\lambda,\mu }=0.
\eeq
These brackets were generalized for the higher order operators $L$ in
\cite{Ad}.

The recurrence operator $B_1B_0^{-1}=C$ generates all the right-hand sides
of all higher KdV systems:
\beq
C(0)=u_x,\ C^2(0)=6uu_x-u_{xxx},\ldots,\ C^n(0)=
\partial_x\left(\frac{\delta I_{n-1}}{\delta u(x)}\right). \label{ROp}
\eeq
It gives also a simple proof of a very useful identity (\cite{GD1}):
\beq
\int \frac{\delta I_n}{\delta u(x)}dx=(const) I_{n-1}  .\label{GD1}
\eeq
All these identities are local and can be used for the class of periodic
functions as well. However, we shall see in the next section that the direct
analog of GGKM procedure does not lead to the integration procedure. We are
going to use a different approach.

\section{Spectral theory of periodic Schr\"odinger operators.
Finite-gap potentials}

The spectral theory for the periodic potentials on the whole line $x$
is based on the monodromy matrix as in case of the
scattering theory. However, in the periodic case with a period $T<\infty$,
we have nothing like the selected point $x=\infty$ for the definition of the
monodromy matrix (as it was for the rapidly decreasing case $(T=\infty)$--see
Section 2 above). Any point $x_0$ can be used. Let us fix initial point $x_0$
and choose a special basis of the solutions $C(x,x_0,\epsilon),
\ S(x,x_0,\epsilon)$ for the spectral equation
$LC=\epsilon C,\ LS=\epsilon S$ such that  for $x=x_0$ we
have:
\begin{eqnarray} \left ( \begin{array}{cc} C&S\\C_x&S_x \end{array}
\right )= \left (\begin{array}{cc} 1&0\\0&1 \end{array} \right )\label{Basis}
\end{eqnarray}
\noindent
The shift operator $T:x\to x+T$ in the basis $C,S$
defines the monodromy matrix
\beq
\widehat T(x_0,\epsilon)=\left ( \begin{array}{cc} a&b\\c&d \end{array} \right ) ,
\label{Mon1}
\eeq
\beq
C(x+T)=a\ C(x)+b\ S(x),S(x+T)=c\ C(x)+d\ S(x) .
\eeq
The key element of the periodic spectral theory is a notion of the, so-called,
Bloch waves or Bloch-Floquet eigenfunctions.
We present here some essential properties of these functions
without any proofs. An exposition of this theory may be found
in the Encyclopedia article
\cite{DKN1}, where the main ideas of the proofs are
clearly presented for the difference Schr\"odinger operator
(it is much simplier).

By definition, the Bloch-Floquet functions are
solutions of the Schr\"odinger equation that  are at the same time
eigenfunctions of the shift operator, i.e.
\beq
L\psi=\epsilon\psi, \ T\psi(x)=\psi(x+T,\epsilon)=
\exp (\pm ip(\epsilon )T)\psi (x).\label{PSI}
\eeq
We uniquely normalize $\psi$ by the condition $\psi|_{x=x_0}=1$.

For any complex number $\epsilon$ the eigenvalues,
$w_{\pm}(\epsilon)=\exp(\pm p(\epsilon)T)$, of the shift operator are defined
by the characteristic equation for the monodromy matrix.
>From the Wronskian property it follows that $\det \widehat T=1$. Therefore,
the characteristic equation has the form
\beq
w^2-({\rm tr}\  \widehat T)w+1=0. \label{char}
\eeq
The multivalued function $p(\varepsilon)$ is called quasi-momentum.

\begin{figure}[th] 
\footnotesize
$$
\epsfbox{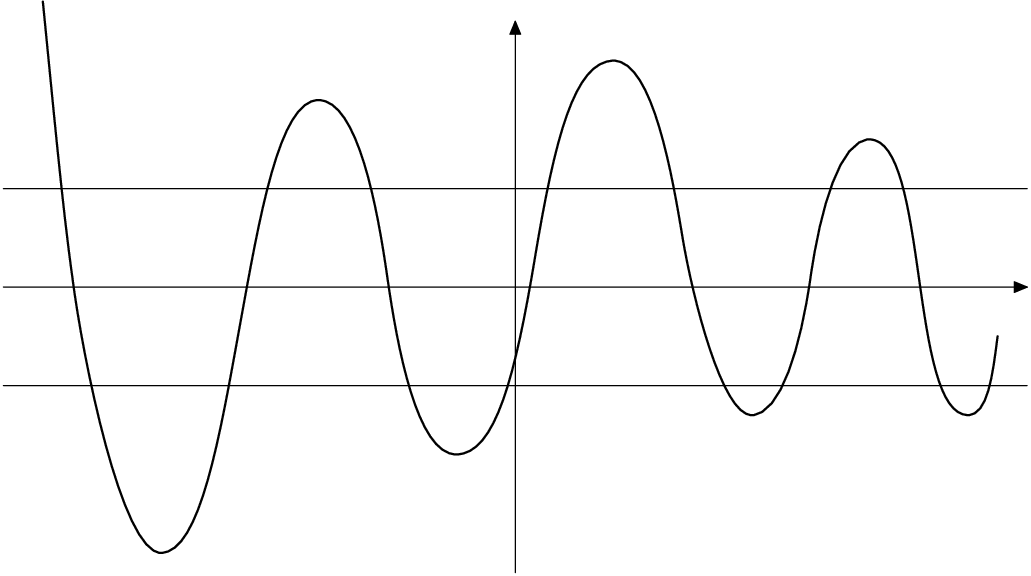}
$$
\caption{}
\end{figure}
\vspace{-2mm}

The spectrum of the Schr\"odinger operator on the whole line is a union
of spectral zones which are segments of the
real line of the variable $\epsilon$, where the quasi-momentum is real.
The latter condition is equivalent to the inequality:
\beq
|{\rm Tr} \ T|=2\cos (pT)\leq 2. \label{Mon4}
\eeq
The typical graph of the function $f(\epsilon)=\cos (pT)$
is presented on the fig\,1. In particular, its extreme points
$f'(\epsilon)=0$ ''generically'' are located inside of the gaps  (i.e. for
the open and everywhere dense set of periodic potentials we have $|f|>1$
at the extreme points, and there is only one extremal point in each gap).
For some special cases we may have $f=\pm 1$ at the extreme point.
Such point lies inside the spectral zone. However,  generic
perturbations create new small gap nearby this point --- see fig\,2.
This point is a double point of periodic or anti-periodic spectrum with
the boundary conditions $\psi (x)=\pm \psi(x+T)$.

\begin{figure}[h]
\footnotesize
$$
\epsfbox{figure2.ps}
$$
\caption{}
\end{figure}

The Riemann surface of the Bloch-Floquet functions is defined by the equation
\beq
z^2= \cos (p(\epsilon) T ), \label{RimS4}
\eeq
\noindent but this surface is nonsingular
only for the generic case (when there are no double points of the
periodic or anti-periodic problem for the Schr\"odinger operator).
For the large $\epsilon\to +\infty$ the following asymptotics
is valid for the gaps (see fig.\,3).

1. The length of gaps tends to zero with its rate depending
on the smoothness of the potential (this rate is exponential for the analytic
potentials).

2. All gaps are located nearby the points $\epsilon_m=4\pi^2 m^2 T^{-2}$,
and the distance of gap from this point tends to zero.

A nonsingular Riemann surface of the Bloch-Floquet solutions is defined
with the help of the equation
\beq
z^2=(\epsilon-\epsilon_0)(\epsilon-\epsilon_1)\ldots
(\epsilon-\epsilon_n)\ldots , \label{RimS}
\eeq
where $\epsilon_0< \epsilon_1< \ldots,$ are simple eigenvalues of the
periodic and anti-periodic spectral problems.

\begin{figure}[h]
\footnotesize 
$$
\epsfbox{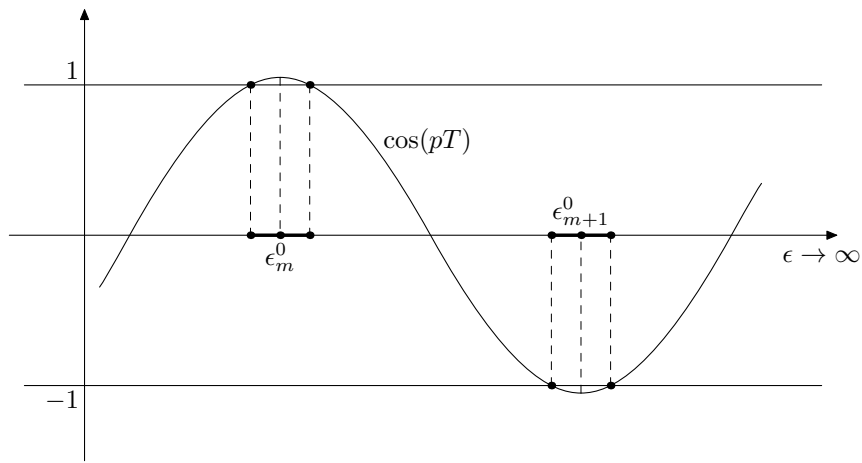}
$$
\begin{quote}
\caption{
For the generic $T$-periodic potentials the gaps are 
located near the points $\epsilon_m=4\pi m^2 T^{-2}$ for 
$\epsilon\to\infty$. Here $m$ are the integer numbers.}
\end{quote}
\end{figure}
\vspace{-2mm}

As it follows from the Lax representation for all higher $KdV_n$ systems
(\ref{LR2}), these boundary spectral points for the
periodic operator $L$  in the Hilbert space $L^2(R)$ of the square integrable
complex-valued functions on $x$--line are integrals of motion for
the KdV hierarchy.

We can say that the Riemann surface of the Bloch-Floquet solutions as a whole
is an integral of motion for the KdV hierarchy.
How many potentials correspond to the same Riemann surface
(i.e. have the same spectrum in $L^2(R)$)?

The original idea to introduce a special class of potentials for which this
problem can be solved effectively was proposed in \cite{N} and is based on the
KdV hierarchy. This idea naturally combines two different ways to
describe the corresponding isospectral
manifold of potentials. Both are in fact closely related to each other.

\noindent{\bf The first approach -- use of the KdV hierarchy and Hamiltonian
dynamics}:

Let us consider a stationary equation for some linear combination of higher
$KdV_n$ flows (\ref{LR2}). It is an ordinary differential equation
that can be written in the form:
\beq
\delta (I_n + c_1I_{n-1}+\ldots +c_nI_0 +
c_{n+1}I_{-1})=0. \label{KdVSt}
\eeq
This is a finite-dimensional Hamiltonian system with $n$ degrees of freedom
depending on $n+1$ parameters $c_1,\ldots ,c_{n+1}$. It is a completely
integrable Hamiltonian system because there is a pencil of the commuting flows.
This pencil coincides with the restriction of all higher $KdV_m$ on this
stationary subset of functions given by the equation (\ref{KdVSt}).
Therefore, its generic nonsingular solution is expected to be a periodic or
quasiperiodic function of $x$.

In the next section we shall construct for (\ref{KdVSt}) some kind of the Lax
representation
\beq \Lambda_x=[Q(x,\lambda ),\Lambda (x,\lambda )], \ \
\Lambda=\Lambda_n+\sum_{k=1}^{n+1} c_k\Lambda_k ,
\label{LR3}
\eeq
\noindent using $2\times 2$ traceless matrices depending on the parameter
$\lambda$ and $(u,u_x,\ldots) $  polynomially.
In the today  terminology some people call them {\it loop groups}.
Note that in all soliton systems this
$\lambda$-loops have very specific $\lambda$-dependence
(polynomial, rational and in some exotic examples -- elliptic functions).

Lax representation (\ref{LR3}) implies that an algebraic Riemann surface defined
by characteristic equation
\beq
\det (\Lambda (\lambda )-zI)=P(\lambda,z)=0,
\label{RimS1}
\eeq
does not depend on $x$ and therefore is an integral of
(\ref{KdVSt}). The same Riemann surface can be extracted from
the {\it commutativity equation} $[L,A]=0$, which according to (\ref{LR2})
is equivalent to (\ref{KdVSt}).

It should be emphasized that the latter form of (\ref{KdVSt}), i.e.
the commutativity condition for two ordinary differential operators
\beq
[L,A_n+\sum c_kA_{n-k}]=[L,A]=0, \ L=-\partial_x^2+u, \label{CE}
\eeq
\noindent
was considered formally (i.e. locally in the variable $x$
without any periodicity assumptions) as a pure algebraic
problem in 1920-s (see \cite{BC, BC1}). Even the formal algebraic Riemann
surface (\ref{RimS1}) appeared as a relation $P(L,A)=0$.

According to our logic however, the corresponding system of equations
in the variable $x$ is Hamiltonian and completely integrable.
Therefore, its generic solution is quasiperiodic in $x$, containing a dense
family of periodic solutions.  We may ask about the spectrum of the
corresponding  operators $L$ in $L^2(R)$ and boundaries of gaps.
Remarkably, they exactly coincide
with the branching points of the Riemann surface defined by equation
(\ref{RimS1}). Therefore, for the periodic potential which satisfy
equation (\ref{KdVSt}) the nonsingular spectral Riemann surface
of the Bloch-Floquet solutions is an algebraic Riemann surface of genus $g=n$.
The spectrum of such an operator contains only a finite number of gaps
$[\epsilon_{2j-1},\epsilon_{2j}],\ j=1,\ldots,n.$

This key step unifies the first approach with the second one (below).  So the
solution of the inverse spectral problem can be identified with the process of
solution of some special families of completely integrable systems using
Riemann surfaces.

\noindent{\bf The second approach -- periodic spectral theory in the Hilbert
space $L^2(R)$}:

For any periodic Schr\"odinger operator $L=-\partial^2+u(x=t_0,t_1,\ldots )$
we have already defined the Riemann surface $\Gamma$ of Bloch-Floquet solutions
with the help of the monodromy matrix $\widehat T$.
For the finite-gap potentials a graph of the function $f=\cos(p(\epsilon)T=
1/2{\rm tr}\  \widehat T$ is highly degenerate (see fig.2 and compare with
fig.1). For all real and large enough $\epsilon$ we have $|f|\leq 1$.

\begin{figure}[t] 
\footnotesize
$$
\epsfbox{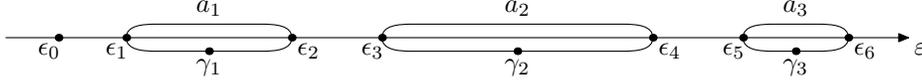}
$$
\begin{quote}
\caption{
The endpoints of gaps $\epsilon_j$ are the branching points of the 
$2$-sheeted Riemann surface. The poles $\gamma_j$ are located in 
the gaps on one branch only. (Here $n=3$)}
\end{quote}
\end{figure}
\vspace{-2mm}

Once again, we ask how to describe all the potentials with the same
finite-gap spectrum and what are additional variables which uniquely define
a finite-gap potential? It turns out that these additional spectral data
are the poles of the Bloch-Floquet functions. It can by shown that
the Bloch-Floquet solutions defined by (\ref{PSI}) and normalized by the
condition $\psi|_{x=x_0}=1$ have exactly one simple pole $\gamma_j(x_0)$
inside each gap or on its boundary. A point of the hyperelliptic
surface $\Gamma$ (\ref{RimS}) can be represented as the complex number
$\epsilon$
and a sign (or a branch) of the radical
$z=\sqrt{(\epsilon-\epsilon_0)(\epsilon-\epsilon_1)\cdots}$.
The branches of the radical coincide at the boundaries of gaps.
Therefore, each gap should be considered as a closed cycle
$a_j,j=1,2,\ldots n$ (see fig.\,4). Any pole of
$\psi$ should be considered as  a point $\gamma_j$ on the
cycle  $a_j$. The above-mentioned statement that there is only one
pole in a gap means, in particular, that
$\psi$ has  no pole at the point $(\gamma_j, +)$ if it already has pole
at $(\gamma_j,-)$ and vice versa.
Geometrically, the total set of poles $(\gamma_1,\ldots,\gamma_n)$
represents a point on the real torus $T^g=a_1\times \ldots \times a_n$.
As we shall see later, this set of poles completely determines the original
potential. Different points of the real torus lead to the different potentials
(if normalization point $x_0$ is fixed).

\section{Periodic analog of GGKM.
Zero Curvature representation for the KdV hierarchy
and corollaries}

As it was emphasized above, the definition of the monodromy matrix
in the periodic case  depends on a choice of initial point $x_0$.
For different choices of the initial point the corresponding monodromy
matrices are conjugated (because they represent the same linear transformation
in different bases).  Therefore, a dependence of the monodromy
matrix with respect to the choice of the initial point $x_0$ can be described
by the equation:
\beq
T_{x_0}=[Q(\epsilon,x_0),T], \ Q=\left
(\begin{array}{cc}0&1\\\epsilon-u(x_0)&0 \end{array}\right ).\label{Mon2}
\eeq

\noindent In the same way, for the isospectral deformations corresponding
to the KdV hierarchy (i.e, all higher $KdV_n$ systems) we can establish
the following equations (the periodic analogs of GGKM):
\beq
T_{t_j}=[\Lambda_j,T], \ \Lambda_0=Q, \ t_0=x_0 .\label{Mon3}
\eeq
A compatibility condition of equations (\ref{Mon3}) for any pair of
variables
$t_i,t_j$ implies the following
{\it zero curvature representation} for the KdV hierarchy
(where periodic boundary conditions are already inessential)
\beq
[\partial_{t_i}-\Lambda_i,\partial_{t_j}-\Lambda_j]=0.\label{ZC}
\eeq
Using the Lax representations (\ref{LR2}) for all higher $KdV_n$, we can
express all  the matrices $\Lambda_j$ as polynomials in the variable
$\epsilon$ and variables  $u(x_0),u_{x_0}(x_0),\ldots $.
For example, for the ordinary KdV we have
(replacing $x_0$ by $x$):
\beq
\Lambda_1=\left (\begin{array}{cc}-u_x&2u+4\epsilon\\
-4\epsilon^2+2\epsilon u+2u^2-u_{xx}&u_x \end{array} \right )\label{KdV2}
\eeq
The matrices $\Lambda_k$ can be completely reconstructed
from equations (\ref{ZC}) and from the following
properties of matrix elements:
\beq
\Lambda_k=(const)\left (\begin{array}{cc}a_k&b_k\\c_k&d_k \end{array}
\right ),\label{Mat}
\eeq
\beq
d_k+a_k={\rm Tr} \Lambda_k=0, \  b_k=\epsilon^k+u/2\epsilon^{k-1}+\ldots,
\eeq
\beq
\det \Lambda_k=-(a^2+bc)=(const)R_{2k+1}(\epsilon)=(const)(\epsilon^{2k+1}+
\ldots ),
\eeq
\beq
2a_k=-b_{kx}, (R_{2k+1})_x=-b_k.
\eeq
For $i=0$, equations (\ref{ZC}), where $t_0=x$, give the zero-curvature
representation for the $KdV_n$ system:
\beq
\p_x\Lambda_n-\partial_{t_n}Q=[Q,\Lambda_n].
\label{ZC1}
\eeq
\noindent From this representation and from the periodic analog of GGKM
we are coming to the following results for the stationary
higher $KdV_n$ system (\ref{LR3}):

1. The monodromy matrix $T$ commutes with $\Lambda$
\begin{eqnarray} [T,\Lambda]=0. \label{Com2}
\end{eqnarray}
\noindent Therefore, they have common eigenvectors. It implies, in particular,
that the Bloch-Floquet function $\psi_{\pm}$ is single-valued on the algebraic
Riemann surface associated with the matrix $\Lambda$;

2. The stationary higher KdV admits an $\epsilon$-parametric Lax-type
representation in the variable $x$ (\ref{LR3}):
\begin{eqnarray}
(\Lambda)_x=[Q,\Lambda ],\ \Lambda=\Lambda_n+\sum c_k\Lambda_k \ .
\label{LR4}
\end{eqnarray}
\noindent Therefore, we have a full set of conservation laws organized in
form of the Riemann surface (i.e.
all coefficients of the polynomial $P$ are $x$-independent):
\beq
\det (\Lambda(\epsilon )-zI)=P(\epsilon,z)=0, \label{Rs}
\eeq
\beq
P(\epsilon,z)=z^2-R_{2n+1}(\epsilon)=(const)(\epsilon-\epsilon_0)\ldots
(\epsilon-\epsilon_{2n})\label{RimS3}.
\eeq
\noindent These points $\epsilon_j$ are exactly the boundaries of
gaps for periodic potentials because the Riemann surfaces of the matrices
$T$ and $\Lambda$ coincide. The matrix element $b_{12}$ for the matrix
$\Lambda$ determines another set of points
\begin{eqnarray}
b_{12}(\epsilon)=(const)(\epsilon-\gamma_1)\ldots (\epsilon-\gamma_n).
\label{Points}
\end{eqnarray}
\noindent These points  coincide with projections of the zeroes of Bloch
wave $\psi_{\pm}$ as the functions of the variable $x$ or poles as the
functions of $x_0$ (see the next section).

An original approach to the solution of the inverse spectral problem was
based on the use of the following trace type formula for the potential:
\beq\label{U}
-u(x)/2+const=\gamma_1(x)+\ldots \gamma_n(x).
\eeq
>From (\ref{LR4}) and (\ref{Points}),  ``Dubrovin equations''
defining dynamics in $x$ of the points $\gamma_j$ can be derived. They have
the form
\beq \label{DE}
\gamma_{jx}=
\frac{\sqrt{R_{2n+1}}(\gamma_j)}{\prod_{k\neq j}(\gamma_j-\gamma_k)}
\eeq
These equations (see \cite{D,DMN}) can be
linearized by the, so-called, Abel transformation.

Consider the first kind differentials on the Riemann surface
(i.e. holomorphic 1-forms without poles
anywhere - even at the infinity). The basic first kind forms are
\begin{eqnarray}\label{HForms}
\omega_j=\frac{\epsilon^{j-1}d\epsilon}{\sqrt{R_{2n+1}(\epsilon)}},
\ j=1,\ldots, n.
\end{eqnarray}
It it is convenient to choose a normalized basis taking linear combinations
$\Omega_s=\sum_jv_{sj}\omega_j$ such
that
$$\oint_{a_j}\Omega_s=\delta_{sj},$$
for all the gaps $a_j$.

Fix a set of paths  $\kappa_j$ from the point $P_0=\infty$ to the points $P_j$
on the Riemann surface.  Abel transformation is defined by the formula
\begin{eqnarray}\label{Abel}
A_p(P_1,\ldots ,P_n)=\sum_j \int_{\kappa_j}\Omega_p\ .
\end{eqnarray}
Equations (\ref{DE}) after the Abel transform become linear:
\beq\label{Angles}
A_{px}=U_q={1\over 2\pi}\oint_{b_q}dp\ .
\eeq
Here the closed paths $b_q$ are ''canonically conjugated'' to the paths
$a_j$ (see fig.\,5). It means that the intersection numbers are 
$a_j\circ b_q=\delta_{jq}$.
The differential $dp$ is equal to $p_{\epsilon}d\epsilon$
where $p(\epsilon )$ is a multi-valued quasimomentum. It is a
second kind differential (i.e. meromorphic 1-form on the surface $\Gamma$
with a pole of order 2 at the point $\infty$ with negative part
$dk=-(dw)w^{-2}$ in the local coordinate $w=k^{-1}$ near $\infty$),
and therefore has the form
\beq\label{Quasi}
dp=\frac{(\epsilon^{n+1}+\sum_{j>0}
v_j\epsilon^{n-1-j})d\epsilon}{\sqrt{R_{2n+1}(\epsilon)}}\ .
\eeq
All the coefficients $v_j$ can be found from the
normalization condition:
\beq
\oint_{a_j}dp=0,\ j=1,\ldots, n\ .     \label{Norm}
\eeq
Following the classical XIX century theory of Riemann surfaces, formulas
(\ref{U}-\ref{Norm}) lead  to some expression of the
potential through the $\theta$-functions avoiding the calculation of the
eigenfunction $\psi_{\pm}$ (see \cite{DN,D,IM,MvM}).
The most beautiful $\theta$-functional formula for the potential was
obtained in the work \cite{IM}.

\begin{figure}[t] 
\footnotesize
$$
\epsfbox{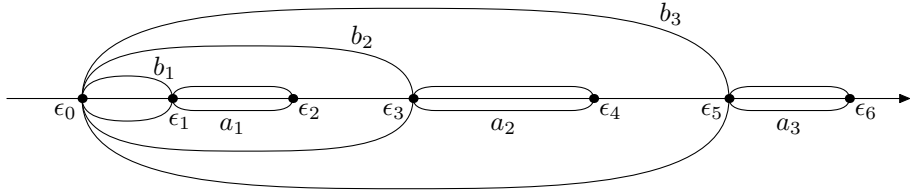}
$$
\begin{quote}
\caption{
The canonical basis of cycles $(a_i,b_j)$ on the Riemann surface~$\Gamma$
with branching points $\epsilon_0,\dots,\epsilon_{2n}$. (Here $n=3$)}
\end{quote}
\end{figure}

However, in the next section we shall not follow this approach
for solving the inverse problem which has been outlined above, but use
another approach proposed for the first time in the article
\cite{DMN} (see appendix, based on  the idea of A.Its):
it is possible to calculate all the family of
the Bloch-Floquet functions. The original approach has been  used
later in the problems, where the effective calculation
of eigenfunction is impossible (like in the cases of {\it higher rank}
commuting OD linear operators (see for example \cite{KN}).

\section{Solution of the periodic and quasiperiodic inverse spectral problems.
Baker-Akhiezer functions}

We are going to solve the inverse spectral problem for the finite-gap
potentials following the scheme proposed in \cite{K,Kf,K1} and based on the concept
of the Baker-Akhiezer functions.
These functions are uniquely defined by their analytical
properties on the spectral Riemann surface. As we shall see later
this scheme is evenly applicable to the solution of inverse problems
in two-dimensional case where the corresponding analytical properties
naturely generalize the analytical properties of the Bloch-Floquet
solutions for finite-gap Schr\"odinger operators.

Let us start with the following real inverse spectral data:

1. Riemann surface $\Gamma$ given in the form
$$z^2= (\epsilon-\epsilon_0)\ldots (\epsilon-\epsilon_{2n}),$$
where all numbers $\epsilon_j$ are real;

2. a set of $n$ real points $\gamma_j\in \Gamma$ such that
there is exactly one point $\gamma_j$ on the cycle $a_j$.

Below we shall for brevity identify the point $\gamma_j$ as
a point on $\Gamma$ with its projection on complex $\epsilon$--plane.
The condition that there is only one point on each cycle means that
projection of the points satisfy the restriction:
\beq
\epsilon_{2j-1}\leq \gamma_j\leq \epsilon_{2j},\ j=1,\ldots n,
\eeq
By definition, the complex inverse spectral data are the same data where
the Riemann surface (i.e.  its branching points $\epsilon_k$) and the points
$\gamma_j\in\Gamma$  are  arbitrary complex points.

As we shall see below a generic set of algebraic-geometric
spectral data leads to
the explicit solution of the inverse spectral problem in terms of the
Riemann $theta$-function. The corresponding potentials are complex meromorphic
{\it quasi-periodic} functions of the variable $x$.
For the real data described above, we are coming to the smooth
(even analytic) quasiperiodic potentials
with their periods expressed through the hyperelliptic integrals (see below).
It is necessary to mention that there is no way to find simple and effective
criteria for the potential to be periodic in terms of these data.
The periods depend on the Riemann surface, only.
Of course, we may write the condition
that all the corresponding hyperelliptic integrals are commensurable,
however this condition in transcendental. Recently, based on the results of
\cite{kr_tau,Gr}, the effective numerical approach has been
developed for the solution of this problem. It is based on the discovery of
some specific dynamic
systems on the set of potentials which preserve all the periods but change
spectrum. Following \cite{Gr}, we start from one periodic potential
and create all others using these dynamic systems.

In the last section we shall present some examples of finite-gap potentials
written in terms of elliptic functions. They are periodic in the variable
$x$ (even double-periodic as functions of the complex variable $x$). The first nontrivial examples
different from the classical Lame potentials $u(x)=n(n+1)\wp(x)$ were
found in \cite{DN}. This subject was developed in the works
\cite{kr1,bab,zab,wig,VT}.

We define the Baker-Akhiezer function $\psi=\psi(x,t_1,\ldots t_n;P)$
for the parameters $x,t_j$ and the point $P=(\epsilon,\pm)\in\Gamma,$
by its analytical properties on $\Gamma$ with respect to the variable $P$.
For the case $t_1=t_2=\cdots=0$ these analytical properties are just the
same as the analytical properties of the Bloch-Floquet solutions of the
periodic finite-gap operator.

>From pure algebraic-geometric arguments it follows that there exists
a unique  function $\psi$  such that

1. it is meromorphic on $\Gamma$ outside the infinity and has at most
simple poles at the points $\gamma_j,\ j=1,\ldots,n$;

2. in a neighborhood of the infinity the function $\psi$ has the form
$$\psi=\exp [xk+t_1k^3+\ldots +k^{2n+1}t_n](1+\xi_1(x,t)k^{-1}+...)\ ,$$
\noindent
where $k^2=\epsilon$ and therefore, $k^{-1}$ is a local coordinate on the
Riemann surface near infinity.

A proof of this statement is identical to the proof of the existence
and uniqueness of general Baker-Akhiezer functions that were
introduced in \cite{K} for construction of exact solutions of two-dimensional
KP equation and all associated Zakharov-Shabat hierarchies.

General Baker-Akhiezer function is defined with the help of an arbitrary
Riemann surface of finite genus $n$ instead of special hyperelliptic
(i.e. two-sheeted) surfaces. We fix an arbitrary "infinity" point $P_0$
on it; local coordinate $k^{-1}=w$ where $w(P_0)=0$ ; generic set of points
$(\gamma_1,\ldots, \gamma_n)$, and numbers
$(x=\tau_1,\tau_2,\ldots ,\tau_k,\ldots) $.
The corresponding Baker-Akhiezer function has the same analytical properties
as above but, (2) is replaced by the following:
\beq
\psi=\exp [xk+\tau_1k^2+\tau_2k^3+\ldots ](1+\xi_1(x,\tau)k^{-1}+\ldots)\ .
\label{Ps}
\eeq

\noindent
For the KdV hierarchy we have a hyperelliptic Riemann surface,
$k^2=\epsilon$, and $\tau_{2k}=0,\tau_{2j+1}=t_j$.

The use of such type of functions (dependent only on a single parameter $x$)
has been proposed by Baker in his note \cite{B}, for the
common eigenfunctions of two commuting
OD linear operators. He expected that this construction will
improve the results of \cite{BC,BC1}. He made also very interesting
conjecture that this approach may seriously improve the classical theory
of $\theta$-functions. Unfortunately, this program was not realized
and was forgotten.
The soliton theory appeared many decades later absolutely independently.
In the 70-s it started to use  such kind of functions on Riemann surfaces
in the process of  solution of a periodic problem for the  KdV type systems
and inverse spectral periodic problems. In the classical spectral theory
Akhiezer (\cite{Ak}) was the first to use some special cases of this function
for the construction   of
some examples of operators on the half-line $x\geq 0$
with interesting spectral  properties. No one of the authors before the 70-s
had associated anything like that  with periodic problems.
Indeed, this type of functional construction on the Riemann surface
was extracted in 1974 from the work \cite{Ak}.

We shall prove the existence and uniqueness of the Baker-Akhiezer function
and present its exact expression through the $\theta$-functions later.
At this moment we would like to show how the uniqueness of this function
leads to the proof that $\psi$ is an eigenfunction for the
Schr\"odinger operator for our special case. Apply twice the operator
$\partial=\partial_x$ to this function and use analytical properties. After
elementary calculation, we are coming to the formulas:
\begin{eqnarray}
\partial\psi&=&k\psi+\exp[kx+\ldots ](\xi_{1x}k^{-1}+\ldots  ),\nonumber\\
\partial^2\psi&=&k^2\psi+2\xi_{1x}\psi+O(k^{-1})\exp [kx+\ldots ].
\end{eqnarray}
>From the latter equations, we get the equality
$$(\partial^2-k^2-2\xi_{1x})\psi=O(k^{-1})\exp [kx+\ldots ].$$

\noindent
The left-hand side is globally well-defined function on the
same Riemann surface because $k^2=\epsilon$. It has the same poles
independent of parameters. Up to the same exponential factor, it is of the
order $O(k^{-1})$ at the infinity.  Therefore,
it is equal to zero due to the uniqueness of the
Baker-Akhiezer function.

So we are coming to the conclusion that
\begin{eqnarray}
L\psi=\epsilon\psi,\ L=-\partial^2+u, \ u=-2\xi_{1x}.
\end{eqnarray}
We can apply the operators
$\partial_j=\partial_{t_j}$ to the function $\psi$:
\begin{eqnarray}
\partial_j\psi=k^{2j+1}\psi+\exp [kx+\ldots ]
(\xi_{1j}k^{-1}+\xi_{2j}k^{-2}) .
\end{eqnarray}
As in the previous case, we can easily construct a linear operator
$A_j=\partial_x^{2j+1}+\ldots $ with the coefficients independent of $k$
such that $(\partial_j-A_n)\psi=O(k^{-1})\exp (kx+\ldots )$. Using global
analytical properties on the Riemann surface, we deduce from this that
the left-hand side is equal to zero as before.
The compatibility condition of these pairs of equations
is exactly a $KdV_n$-system for the potential $u(x,t_1,\ldots,t_m)$.

Following the works \cite{K,Kf,K1}, we can prove in the same way that
for the general Baker-Akhiezer function $\psi(x,\tau,P)$ associated with
arbitrary Riemann surface with fixed local coordinate near
a puncture, the following equations are valid:
\begin{eqnarray}
(\partial_{\tau_k}-L_k)\psi=0, \ k=2,3,\ldots. \label{Lk}
\end{eqnarray}
\noindent
Here $L_k=\partial_x^k+\ldots $
are linear OD operators acting on the variable $x$
with coefficients depending on parameters $\tau$. These coefficients are
differential polynomials in the coefficients of the expansion of the regular
factor of the Baker-Akhiezer function at the puncture (infinity).
They are uniquely defined by the condition that the expansion of the
left-hand side of (\ref{Lk}) at the puncture has the form
$O(k^{-1})\psi$.

The compatibility conditions
\begin{eqnarray}
[\partial_{\tau_j}-L_j,\partial_{\tau_k}-L_k]=0. \label{zk}
\end{eqnarray}
are equivalent to non-linear partial differential equations for the
coefficients of the operators $L_k$.
For the case $j=2,k=3, \tau_2=y,\tau_3=z$,  the operators $L_2$ and $L_3$
have the form
\beq L_2=\p_x^2-u(x,y,t), \ L_3=\p_x^3-{3\over 2}u\p_x+w(x,y,t) ,\label{kp}
\eeq
where we consider the dependence of the coefficients with respect to the
first three variables $\tau_1=x,\tau_2=y,\tau_3=t$, only .
The coefficient $u(x,y,t)$ is equal to $u=2\xi_{1x}(x,y,t)$, where
$\xi_1$ is the first coefficient of the expansion (\ref{Ps}).

>From (\ref{zk}) we get a system of two equations for two coefficients $u(x,y,t)$ and
$w(x,y,t)$ which can be reduced for an equation for $u(x,y,t)$.
The reduced equation is the famous KP equation
\beq
{3}u_{yy}=(4u_t-6uu_x+u_{xxx})_x. \label{kp2}
\eeq
It appeared in physics literature for the investigation of the
transversal stability of the  KdV solitons (see \cite{KP}) and is one of the
most natural physical two-dimensional analogs of KdV.
Lax representation for it was found in \cite{ZS1,Dr}.
We call the whole set of higher systems (\ref{zk}) the KP hierarchy.
We shall discuss the periodic problem for this equation at greater length
in the next section.

For the special choice of the hyperelliptic Riemann surface
and $k=\pm\sqrt{\epsilon}$ we have $k^{2s}=\epsilon^s$, where the
function $\epsilon$ is well-defined globally as a meromorphic function on the
Riemann surface $\Gamma$. Therefore, we may represent globally the
corresponding Baker-Akhiezer function in the form
\begin{eqnarray}
\psi=\tilde{\psi}\exp (\tau_2k^2+\tau_4k^4+\ldots),
\end{eqnarray}
\noindent
where  $\tilde{\psi}$ does not depend on the parameters $\tau_{2j}$.
So in this case all the KP hierarchy reduces to the KdV hierarchy.

The relationship of this constructions with commuting OD linear
operators is as follows. Let $f(P)$ be a meromorphic function on the
Riemann surface $\Gamma$ with one pole at the puncture $P_0$.
Its negative part written in the parameter $k$  is some polynomial
$q(k)=q_1^l+q_2k^{l-1}+\ldots +q_lk$.
Apply the operator $A_f=q_1L_l+q_2L_{l-1}+\ldots +q_1\partial_x$
to the function $\psi$. By the definition of the operators $L_k$,
we can see that
\begin{eqnarray}
(A_f-f)\psi=O(k^{-1})\psi.
\end{eqnarray}
\noindent
We conclude as before, that the difference $A_f\psi-f(P)\psi=0$.

For any pair of functions $f$ and $g$
on the Riemann surface $\Gamma$ with poles at the point
$\infty$, we get a pair of the commuting OD scalar linear operators
$A_f,A_g$ such that $A_fA_g=A_gA_f$. In a special case of the
hyperelliptic Riemann surface
$z^2=(\epsilon-\epsilon_0)\ldots (\epsilon-\epsilon_{2n})$,
we have pair of functions $f=\epsilon,\ g=z$, leading to the Schr\"odinger
operator $L=-\partial_x^2+u$ commuting with  the second operator of order
$2n+1$, because $z=k^{2n+1}+\ldots $

Now we return back to the problem of existence and uniqueness of the
Baker-Akhiezer function. The simplest way to prove this existence is
to define this function by exact formula in terms of the $\theta$-function
and meromorphic differentials. Let us recall first necessary information.

As in the previously discussed case of hyperelliptic curves, we introduce
a basis of cycles $a_j,b_j, j=1,\ldots,g, $ on a Riemann surface $\Gamma$
of genus $g$ with canonical matrix of intersections:
$a_i\ \cdot\  b_j=\delta_{ij}$, and a basis $\omega_i$ of holomorphic
differentials normalized by the condition
\beq
\oint_{a_{i}}\omega_j=\delta_{ij}.
\eeq
The matrix $B=(B_{ij})$ of $b$-periods of these differentials
\beq
B_{ij}=\oint_{b_i}\omega_{j}   ,
\eeq
is symmetric and has positively defined imaginary part. The Riemann
$\theta$-function is a function defined with the help of this matrix by
the formula:
\beq
\theta(z|B)=\sum_{m\in Z^g}e^{2\pi i(z,m)+(Bm,m)}   \label{thet}
\eeq
where $z=(z_1,\ldots,z_g)$ is a complex $g$-dimensional vector, $(m,z)$
stands for a standard scalar product and summation is taken over all
integer vectors $m=(m_1,\ldots,m_g)$.
Theta-function is an entire periodic function of $g$ variables $z_j$ and
has the following monodromy properties with respect to the shifts defined
by vectors $B_k$ which are columns of the matrix of $b$-periods:
\beq
\theta(z+B_k)=e^{-2\pi i z_k-\pi i B_{kk}}\theta(z).
\eeq
The basic vectors $e_k$ and the vectors $B_k$ define a lattice ${\cal L}$
in $C^g$ which determines $g$-dimensional complex torus $J(\Gamma)=C^g/{\cal L}$
called Jacobian of the curve. The Abel map $A:\Gamma\to J(\Gamma)$ is defined
by the formula
\beq
A_k(P)=\int_{P_0}^P\omega_k.
\eeq
Note that the vector $A(P)$ with coordinates $A_k(P)$ depends on the choice
of path of integration but its ambiguity just coincides with shifts by
vectors of the lattice ${\cal L}$.

>From the monodromy properties of $\theta$-function
it follows that zeros of the multivalued function $\theta(A(P)+Z)$
considered as a function on $\Gamma$ are well-defined.
For a generic vector $Z$ this function has exactly $g$ zeros
$(\gamma_1,\ldots,\gamma_g)$. The vector $Z$ can be expressed in terms of
Abel transforms of these points by the formula
\beq
Z=-\sum_{j=1}^g A(\gamma_j)+K,
\eeq where $K$ is a vector of the Riemann constants.

Let us introduce a set of meromorphic differentials
$d\Omega_i$ that are holomorphic on $\Gamma$ outside the puncture where they
have poles of the form
\beq
d\Omega_i=dk^i(1+O(k^{-i-1})),
\eeq
and normalized in a usual way by the condition
\beq
\oint_{a_i}d\Omega_j=0.
\eeq
The Abelian integrals
\beq
\Omega_i(P)=\int^Pd\Omega_i
\eeq
are multivalued functions on $\Gamma$.

Let $U_j$ be a vector with the coordinates
\beq
U_{jk}={1\over 2\pi i}\oint_{b_k}d\Omega_j.
\eeq
Then from the statements presented above, it follows directly that the formula
\beq
\psi(\tau,P)=\exp(\sum_i\tau_i\Omega_i(P))
{\theta((A(P)+\sum_i U_i\tau_i+Z)\theta(Z)\over
\theta(A(P)+Z)\theta(\sum_i U_i\tau_i+Z)}, \label{f1}
\eeq
correctly defines a function on $\Gamma$ which satisfies all the properties
of the Baker-Akhiezer function.

Suppose now that there exists another Baker-Akhiezer function $\psi_1$.
>From the definition the Baker-Akhiezer functions it follows that the ratio
$\psi_1/\psi$ is a meromorphic function on $\Gamma$ which is equal to 1
at the puncture and with only possible
poles at the zeros of the function $\psi$.
According to (\ref{f1})  the zeros of $\psi$ are zeros of the function
$\theta(A(P)+\sum_iU_i\tau_i+Z)$. Therefore, $\psi$ has
$g$ zeros. The simplest form of the Riemann-Roch theorem (which can be
considered as generalization of the Liouville theorem for Riemann surfaces)
implies that a function on $\Gamma$ with at most $g$ poles at a
generic set of points is a constant. Therefore, $\psi_1=\psi$ and
the existence and uniqueness of the Baker-Akhiezer function is proved.

Now according to the previously established formula
$u=2\p_x\xi_1$, in order to get an exact formula for a solution of the KP
hierarchy it is enough to take the first coefficient $\xi_1(\tau)$
of the expansion of the pre-exponential factor in (\ref{Ps}) at the puncture.
Finally we obtain the expression
\beq
u(\tau)=2\p_x^2\log \theta(\sum_i U_i\tau_i+Z)+const
\eeq
for the finite gap solutions of the whole KP hierarchy.

If we consider only the KP equation, we get the formula
\beq
u(x,y,t)=2\p_x^2 \log \theta(Ux+Vy+Wt+Z)+ const, \label{kpf}
\eeq
where we redenote $x=\tau_1, y=\tau_2, t=\tau_3$ and $U=U_1, V=U_2, W=U_3$.

For the case of the hyperelliptic curve the vector $V=0$ and we get the
Its-Matveev formula for the finite-gap solutions of the KdV equation.

The formula (\ref{kpf}) derived in \cite{Kf,K1} has led
to one of the most important pure mathematical applications of the theory
of non-linear integrable systems. This is the solution of the famous
Riemann-Shottky problem.

According  to  the Torrelli theorem,  the matrix of $b$-periods of normalized
holomorphic differentials uniquely defines the corresponding algebraic curve.
The Riemann-Shottky problem is: to describe symmetric matrices with the
positive imaginary part which are the matrices of $b$-periods of normalized
holomorphic differentials on algebraic curves. One of the authors
conjected that the function $u(x,y,t)$ given by (\ref{kpf})
is a solution of the KP-equation iff the matrix $B$ that defines the
theta-function is the matrix of $b$-periods of normalized holomorphic
differentials on an algebraic curve and $U,\ V,\ W$ are vectors of
$b$-periods of corresponding normalized meromorphic differentials with
the only pole at a point of this curve. This conjecture was proved in
\cite{shiot}.

\section{Spectral theory of two-dimensional periodic operators.
KP hierarchy}

A general algebraic-geometric construction of the finite-gap potentials for
the Schr\"odinger operators and for solutions of the KP hierarchy
that was presented in the previous section has been developed extensively
in years. It is applicable for all soliton systems which are equivalent
to various types of compatibility conditions for over-determined systems
of auxiliary linear problems. In its algebraic form it is in some sense
local and is a sort of inverse transform: from a set of algebraic-geometrical
data to solutions of the integrable non-linear partial differential equations
\beq
\{ algebraic-geometrical \  data\} \longmapsto \{ solutions \ of \ NLPDE \}
\label{2}
\eeq
In a generic case the space of algebraic-geometrical data is a union for all
$g$ of the spaces
\beq
\tilde M_{g,N}=\{\Gamma_g,P_{\alpha},k_{\alpha}^{-1}(Q),\gamma_1,\ldots,
\gamma_g\}, \ \alpha = 1,\ldots,N, \label{2a}
\eeq
where $\Gamma_g$ is an algebraic curve of genus $g$ with fixed local
coordinates $k_{\alpha}^{-1}(Q),k_{\alpha}^{-1}(P_{\alpha})=0,$ in
neighborhoods of $N$  punctures $P_{\alpha}$, and $\gamma_1,\ldots,\gamma_g$
are points of $\Gamma_g$ in a general position. (It is to be mentioned that
$\tilde M_{g,N}$ are ``universal" data. For the given non-linear integrable
equation the corresponding subset of data has to be specified.)

A'posteriory it can be shown that these solutions can be expressed in terms
of the corresponding Riemann
theta-functions and are quasi-periodic functions of all  variables. Whithin
this approach it is absolutely impossible to give an answer to  the  basic
question:  "How many algebraic-geometrical solutions are there? And what is
their role in the solution of the periodic Cauchy problem for two-dimensional
equations of the KP type?"

The answer of the corresponding question in lower dimensions is as follows.
For finite dimensional (0+1) systems a typical  Lax representation
has the form
\beq
\p_t U(t,\lambda)=[U(t,\lambda), V(t,\lambda)] ,  \label{7a}
\eeq
where $U(t,\lambda)$ and $V(t,\lambda)$ are matrix functions that are rational
(or sometimes elliptic) functions of the spectral parameter $\lambda$.
In that case {\it all} the general solutions are algebraic-geometrical and
can be represented in terms of the Riemann theta-functions.

For spatial one-dimensional evolution
equations of the KdV type (1+1)-systems)
the existence of direct and inverse spectral transform
allows one to prove (though it is not always the rigorous mathematical
statement) that algebraic-geometrical solutions are dense in the space of all
periodic (in $x$) solutions.

It turns out that the situation for two-dimensional integrable equations is
much more complicated. For one of the real forms of the KP equation that is
called the KP-2 equation and coincides within (\ref{kp2}), the
algebraic-geometrical solutions are dense in the space of all periodic (in $x$
and $y$) solutions  (\cite{Ksp}).  It seems, that the same statement for the
KP-1 equation which can be obtained from (\ref{kp2}) be replacing $y\to iy$
is wrong. One of the most important problems in the theory of
two-dimensional integrable systems which are still unsolved
is: "in  what sense" the KP-1 equation ahich has the operator representation
(\ref{zk}) and for which a wide class of periodic solution was
constructed, is an "{\it non-integrable}" system.

The proof of the integrability of the periodic problem for the
KP-2 equation is based on the spectral Floquet theory of the parabolic operator
\beq
M=\p_y-\p_x^2+u(x,y),  \label{8}
\eeq
with periodic potential $u(x+l_1,y)=u(x,y+l_2)=u(x,y).$ We are going
now to present the most essential points of this theory which was
developed in \cite{Ksp}. It is the natural generalization of the spectral
theory of
the periodic Sturm-Liouville operator. We would like to mention that despite
its application to the theory of non-linear equations and related topics, the
structure of the Riemann surface of Bloch solutions of the corresponding
linear equation that was found in \cite{Ksp} has been used as a starting
point  for an abstract definition of the Riemann surfaces of the
infinite genus (\cite{trub}).

Solutions $\psi(x,y,w_1,w_2)$ of the non-stationary Shr\"odinger
equation
\beq
(\sigma \partial_y-\partial_x^2+u(x,y))\psi(x,y,w_1,w_2)=0 \label{3.7}
\eeq
with a periodic potential $u(x,y)=u(x+a_1,y)=u(x,y+a_2)$ are called
Bloch solutions, if they are eigenfunctions of the monodromy operators,
i.e.
\beq
\psi(x+a_1,y,w_1,w_2)=w_1\psi(x,y,w_1,w_2),
\eeq
\beq
\psi(x,y+a_2,w_1,w_2)=w_2\psi(x,y,w_1,w_2).
\eeq
The Bloch functions will always be assumed to be normalized so that
$\psi(0,0,w_1,w_2)=1$.
The set of pairs $Q=(w_1,w_2)$, for which there exists such a solution is
called the Floquet set and will be denoted by $\Gamma$. The multivalued
functions $p(Q)$ and $E(Q)$ such that
$$w_1=e^{ipa_1}\ ,\ w_2=e^{iEa_2}$$
are called quasi-momentum and quasi-energy, respectively.

The gauge transformation $\psi \to e^{h(y)}\psi$, where $\partial_y h(y)$ is a
periodic function, transfers the solutions of (\ref{3.7}) into solutions of
the same
equation but with another potential $\tilde u=u-\sigma \partial_y h$.
Consequently, the spectral sets corresponding to the potentials $u$ and
$\tilde u$ are isomorphic. Therefore, in what follows we restrict
ourselves to the case of periodic potentials such that
$\int_0^{a_1} u(x,y)dx=0.$

To begin with let us consider as a basic example
the ``free" operator
$M_0=\sigma \p_y-\p_x^2 $
with zero potential $u(x,y)=0$. The Floquet set of this
operator is parameterized by the points of the complex plane of the
variable $k$
$w_1^0=e^{ika_1}\ ,\ w_2^0=e^{-\sigma^{-1}k^2a_2}$,
and the Bloch solutions have the form
$\psi(x,y,k)=e^{ikx-\sigma^{-1}k^2y}.$
The functions $\psi^+(x,y,k)=e^{-ikx+\sigma^{-1}k^2y}$
are Bloch solutions of the formal adjoint operator
$(\sigma\p_y+\p_x^2)\psi^+=0.$

An image of the map $k\in C \longmapsto (w_1^0, w_2^0)$
is the Floquet set for the free operator $M_0$. It is the
Riemann surface with self-intersections. The self-intersections
correspond to the pairs $k\neq k'$ such that
$w_i^0 (k)=w_i^0 (k'),\ \ i=1,2.$
The latter conditions imply the equations
\beq
k-k'=\frac {2\pi N}{a_1}\ ,\
k^2 -(k')^2=\frac {\sigma 2\pi iM}{a_2}\ ,\label{3.16}
\eeq
where $N$ and $M$ are integers. Hence, all the resonant points have the form
\beq
k=k_{N,M}=\frac{\pi N}{a_1}-\frac {\sigma i Ma_1}{Na_2},
\ N\neq0, \ k'=k_{-N,-M}.\label{3.17}
\eeq
The basic idea of the construction of the Riemann surface of Bloch solutions
of the equation (\ref{3.7}) that was proposed in \cite{Ksp} is to consider
(\ref{3.7}) as a perturbation of the free operator,
assuming that the potential $u(x,y)$ is formally small.

For any $k_0\neq k_{N,M}$ it is easy to construct a formal Bloch solution.
It turns out that the corresponding formal series converges and
defines a holomorphic function of $k_0$ for $|k_0|>M$ big enough and
lies outside small neighborhoods of the resonant points. Moreover,
it can be shown that this function can be extended on the Riemann surface
that can be thought as a surface obtained from the complex plane by
some kind of surgery that creates {\it gaps} in places of the resonant points.

More precisely, if $u(x,y)$ is a smooth real potential that has
analytical continuation in some neighborhood of the real values of $x$ and $y$,
then the corresponding Riemann surface of Bloch-Floquet solutions can be
described in the following way.

Let us fix some finite or infinite subset $S$ of integer pairs $(N>0,M)$ .
The set of pairs of complex numbers $\pi=\{p_{s,1},p_{s,2}\}$ where
$s\in S$ would be called ``admissible", if
\beq
{\rm Re}\ p_{s,i}=\frac{\pi N}{a_1}\ ,\
\vert p_{s,i}-k_s \vert=o(\vert k_s \vert ^{-1}),\ i=1,2,\label{3.50}
\eeq
and the intervals
$[p_{s,1},p_{s,2}]$ do not intersect. (Here $k_s,\ s=(N,M)$ are resonant
points.)

Let us define the Riemann surface $\Gamma(\pi)$ for any admissible set $\pi$.
It is obtained from the complex plane of the variable $k$ by cutting it along
the intervals $[p_{s,1},p_{s,2}]$ and $[-\bar p_{s,1},-\bar p_{s,2}]$ and by
sewing after that the left side of the first cut with the right side of the
second cut and vice versa. ( After this surgery for each cut $[p_{s,1},p_{s,2}]$
corresponds to a nontrivial cycle $a_s$ on $\Gamma(\pi)$.)

For any real periodic potential $u(x,y)$ which can be
analytically extended into some neighborhood of the real values $x,y$,
the Bloch solutions of the equation (\ref{3.7}) are
parameterized by points $Q$ of the Riemann surface $\Gamma (\pi)$
corresponding to some admissible set
$\pi$. The function $\psi(x,y,Q)$ which is normalized by the condition
$\psi(0,0,Q)=1$ is meromorphic on $\Gamma$ and has a simple pole $\gamma_s$
on each
cycle $a_s$. If the admissible set $\pi$ contains only a finite number of
pairs, then $\Gamma (\pi)$ has finite genus and is compactified by only
one point $P_1$ ($k=\infty$), in the neighborhood of which the Bloch
function $\psi$ has the form (\ref{Ps}).

The potentials $u$ for which $\Gamma (\pi)$ has finite genus are called
finite-gap. They
coincide with the algebraic-geometrical potentials.
The  direct spectral transform for periodic operators (\ref{8}) allows us to
prove that like in the one-dimensional case the finite-gap potentials are
dense in the space of all periodic smooth functions in two
variables (\cite{Ksp}).

\section{Spectral theory of two-dimensional Schr\"odinger operator
for fixed energy level and two-dimensional Toda lattice}

In this section we discuss a spectral theory of two-dimensional periodic
Sch\"odinger operator. Unlike the one-dimensional case, spectral data
for 2D linear operator are over-determined and therefore for generic
operators there are no nontrivial isospectral flows. As it was noted in
\cite{man}, deformations that preserve spectral data for {\it one fixed}
energy level do exist. An analog of Lax representation for such system
has the form
\beq
H_t=[A,H]+BH, \label{lab}
\eeq
where $H,A,B$ are two dimensional operators with coefficients depending
on $x,y,t$. Equation (\ref{lab}) is equivalent to the condition that
operators $H$ and $(\p_t-A)$ commute on the space of solutions of the equation
$H\psi=0$. Therefore, (\ref{lab}) describes deformations preserving
all the spectral data associated with the zero energy level of the
operator $H$.

It should be mentioned that until the moment when equations (\ref{lab})
were proposed in the framework of the soliton theory the spectral
problem associated with one energy level of two-dimensional periodic
operators had never been considered.

For the first time an inverse algebraic-geometric spectral
problem for two-dimensional Schr\"odinger operator in magnetic
field
\beq
H=(i\p_x-A_x(x,y))^2+(i\p_y-A_y(x,y)^2+u(x,y) \label{h}
\eeq
was formulated and solved in \cite{DKN}.

Consider the Bloch solutions of the equation $H\psi=\epsilon\psi$, which
by definition are solutions that at the same time  are
eigenfuctions for the shifts operators:
\beq \label{sh}
\psi(x+T_1,y)=e^{ip_xT_1}\psi (x,y),\ \psi(x,y+T_2)=e^{ip_yT_2}\psi(x,y).
\eeq
Here $T_1,T_2$ are periods of the operator $H$, i.e. periods of the potential
$u(x,y)$ and periods of the magnetic field $B(x,y)$, which is defined
by the vector potential $(A_x,A_y)$ by the formula $B=\p_yA_x-\p_xA_y$.

Multivalued quantities $p_x,p_y$ are components of the two-dimensional
quasimomentum. For fixed values of $p_x,p_y$ a spectrum of the operator $H$
restricted on the space of functions satisfying (\ref{sh}) is discrete and
defines different branches of dispersion relations
$\epsilon_j(p_x,p_y), j=1,\ldots$
Level lines $\epsilon_j(p_x,p_y)=\epsilon_0$
in the space of variables $p_x,p_y$
define  the, so-called, {\it Fermi curves}. Of course,
in the solid state physics all the considerations were restricted by real
values  of quasi-momentum.

In \cite{DKN} it was suggested to consider operators for which
 a {\it complex} Fermi curve does exist
and is the Riemann surface of finite genus for some energy level $\epsilon_0$.
Moreover, it was assumed
that this curve is compactified by two {\it infinity} points $P_{\pm}$
in neighborhoods of which the corresponding Bloch solutions have the form
\beq
\psi=e^{k_{\pm}(x\pm iy)}
\left(\sum_{s=0}^{\infty}\xi_s^{\pm}(x,y)k_{\pm}^{-s}\right),
\label{inf}
\eeq
where $k_{\pm}^{-1})$ are local coordinates in the neighborhoods of the
punctures  $P_{\pm}$. It was also assumed that outside the punctures
the functions $\psi(x,y,P)$ considered as a function of the variable
$P$ (which is a point of the complex Fermi curve $\Gamma$) is meromorphic
function with $g$ poles independent of the variables $x$ and $y$.

We present here a solution to this inverse spectral problem in a  more general
form which is necessary for construction of exact solutions to the
two-dimensional Toda lattice which has deep connections with the theory
of 2D Schr\"odinger operators. After that we return back to the spectral
problems.

Let $\G$ be a smooth algebraic curve of genus $g$ with fixed local
coordinates $w_{\pm}(P)$ in neighborhoods of the points
$P^{\pm}, \ w_{\pm} (P^{\pm})=0$. Then for each set of
$g$ points $\gamma_1,\ldots,\gamma_{g}$ in general position
there exists a unique function $\psi_n(T,P),
\ T=\{t_{i}^{\pm},\ i=1,\ldots,\infty,\}$ such that:

$1^0.$ The function $\psi_n$ of the variable $P\in \Gamma$ is
meromorphic outside the punctures and has at most simple poles at the points
$\gamma_s$ (if all of them are distinct);

$2^0.$ In a neighborhood of the point $P^{\pm}$ it has the form
\beq
\psi_n(T,P)=w_{\pm}^{\mp n}
\left(\sum_{s=0}^{\infty} \xi_s^{\pm}(n,T)w_{\pm}^s \right)
\exp \left(\sum_{i=1}^{\infty}w_{\pm}^{-i}t_{i}^{\pm}\right),\ \ \
w_{\pm}=w_{\pm}(P),\label{3.2}
\eeq
\beq
\xi_0^{+}(x,T)\equiv \delta_{\alpha j}. \label{3.3}
\eeq
The proof of this statement, as well as the explicit formula
for $\psi_n$ in terms of Riemann theta-functions, is almost identical
to the way of solution of the inverse problem for finite-gap
Schr\"odinger operators discussed in the previous sections.

Let $d\Omega_j^{(\pm)}$ be a unique meromorphic differential
holomorphic on $\G$ outside the point $P^{\pm}$, which has the form
\beq
d\Omega_j^{(\pm)}=d(w_{\pm}^{-j}+O(w_{\pm})),
\eeq
near the point $P^{\pm}$, and normalized by the conditions
$\oint_{a_k}d\Omega_i^{(\pm)}=0.$
It defines a vector $U_j^{(\pm)}$ with  coordinates
\beq U_{jk}^{(\pm)}={1\over 2\pi i} \oint_{b_k} d\Omega_j^{(\pm)}.
\eeq
Further, let us define the normalized differential $d\Omega^{(0)}$, which is
holomorphic outside the points $P^{\pm}$ where it has simple poles
with residues $\pm 1$, respectively.  From Riemann's bilinear relations it
follows that the vector of $b$-periods of  this differential equals
$2\pi iU^{(0)}$ , where
\beq
U^{(0)}=A(P^-)-A(P^+)\ . \label{3.9c}
\eeq
As in the previous case, one can directly check that the function
$\psi_n(T,P)$ given by the formula:
\beq
\psi_n(T,P)={\theta(A(P)+U^{(0)}n+\sum U_i^{(\pm)}t_i^{\pm}+Z)
\theta (A(P^+)+Z)) \over
\theta (A(P)+Z) \theta (A(P^+)+U^{(0)}n+\sum U_i^{(\pm)}t_i^{\pm}+Z) }
e^{(n\Omega^{(0)}(P)+\sum t_i^{\pm}\Omega_i^{(\pm)}(P))}, \label{3.10}
\eeq
\beq \Omega_i^{(\pm)}(P)=\int^P d\Omega_i^{(\pm)}\ ,
\eeq
is well-defined and has all the properties of the Baker-Akhiezer function.

Note, that for $n=0$ and $t_1^{\pm}=x\pm iy, t_i^{\pm}=0, i>1,$ the analytical
properties of this function coincide with the properties that were described
above as analytical properties of the Bloch solutions for finite-gap
two-dimensional Schr\"odinger operators.

>From the uniqueness of the Baker-Akhiezer functions $\psi_n(T,P)$ it follows
that they satisfy the linear equations
\beq
\p_+\psi_n=\psi_{n+1}+v_n\psi_n ,\ \p_-\psi_n=c_n\psi_{n-1},
\ \p_{\pm}={\p \over \p t_1^{\pm}},\label{2d}
\eeq
where
\beq
v_n=\p_+\varphi_n(T), \ \ c_n=e^{\varphi_n(T)-\varphi_{n-1}(T)}, \ \
e^{\varphi_n}=\xi_0^{-}(T),
\eeq
and $\xi_0^-(T)$ is a leading term of the expansion of $\psi_n$ at the
puncture $P^-$. From (\ref{3.10}) we get the formula
\beq
\varphi_n=
\log {\theta(U^{(0)}(n+1)+\sum U_i^{(\pm)}t_i^{\pm}+Z_0)
\over \theta (U^{(0)}n+\sum U_i^{(\pm)}t_i^{\pm}+Z_0) }+const,
\ Z_0=Z+A(P^+).
\eeq
for algebraic-geometric solutions of 2D Toda lattice which was obtained
in \cite{Ktod}.

Note that (\ref{2d}) imply that  $\psi_0$ satisfies the equation
\beq
\p_+\p_- \psi_0+v_0\p_-\psi_0+(c_1+\p_-v_0)\psi_0,
\eeq
which is gauge equivalent to (\ref{h})
and therefore, we do get a solution of the inverse problem that was
introduced above.

The next important step was done in \cite{NV,VN} where
algebraic-geometric spectral
data corresponding to potential $2d$ Schr\"odinger operators
(i.e. operators with zero magnetic field) were found.

Let $\Gamma$ be a smooth genus $g$ algebraic curve  with fixed local
coordinates
$k_{\pm}^{-1}(Q)$,\\
$k_{\pm}^{-1}(P_{\pm})=0,$ in neighborhoods of two
punctures $P_{\pm}$. Let us assume that there exists a holomorphic
involution of the curve $\sigma: \Gamma \longmapsto \Gamma$
such that $P_{\pm}$ are its only fixed points, i.e.
$\sigma (P_{\pm})=P_{\pm}.$
The local parameters are to be ''odd", i.e.
$k_{\pm}(\sigma (Q))=-k_{\pm}(Q). $
The factor-curve will be denoted by $\Gamma_0$. The projection
\beq
\pi :\Gamma \longmapsto \Gamma_0=\Gamma / \sigma \label{3.6}
\eeq
represents $\Gamma$ as a two-sheet covering of $\Gamma_0$ with the
two branch points $P_{\pm}$. In this realization the involution $\sigma$ is
a permutation of the sheets. As there are only two branching
points $g=2g_0$, where $g_0$ is genus of $\Gamma_0$.
Let us consider a meromorphic differential $d\Omega(Q)$ of the third kind on
$\Gamma_0$ with residues $\mp 1$ at the points $P_{\pm}$. The differential
$d\Omega$ has $g$ zeros that will be denoted
by $\hat \gamma_i,\ i=1,\ldots,2g_0=g.$
Let us choose for each $i$ a point $\gamma_i$ on $\Gamma$ such that
\beq
\pi (\gamma_i)=\hat \gamma_i, \ i=1,\ldots,g.
\eeq
In \cite{NV,VN} it was shown that the Baker-Akhiezer function corresponding
to algebraic-geomet\-ric data which  have been just defined satisfies
the equation
\beq
(\partial_+\partial_- + u(x,y))\psi(x,y,Q)=0, \label{3.11}
\eeq
where
\beq
u=-\partial_- \xi_1^+=-\partial_+ \xi_1^{-}, \label{3.12}
\eeq
and $\xi_1^{\pm}=\xi_1^{\pm}(x_+,x_-)$ are the first coefficients in the
expansion (\ref{inf}).

It should emphasized that though general formula (\ref{3.10})
in terms of the Riemann $\theta$-functions is valid
for the Baker-Akhiezer functions corresponding to the potential Shr\"odinger
operator $H$,  in \cite{NV,VN} it was
found another more effective representation in terms of the, so-called, Prim
theta-function.

A space of holomorphic differentials on $\Gamma$ splits into two
$g_0$-dimensional subspaces of even and odd (with respect to the involution
$\sigma$) differentials. A matrix of $b$-periods of odd differentials
defines the function $\theta_{Pr}(z)$ by the same formula (\ref{thet}).
Then
\beq
\psi={\theta_{Pr}(A^{od}(Q)+U^+x_++U^-x_--Z)\theta_{Pr}(A^{od}(Z))
\over \theta_{Pr}(A^{od}(Q)-Z)\theta_{Pr}(U^+x_++U^-x_--Z)}
e^{i(p^+(Q)x_++p^-(Q)x_-)}. \label{304}
\eeq
Here $p^{\pm}(Q)$ are Abelian integrals of the second kind
normalized differential $dp^{\pm}$ on $\Gamma$ that have poles of the
second order at the points $P_{\pm}$, respectively; vectors $2\pi U^{\pm}$
are the vectors of $b$-periods of these differentials.

As it was mentioned above the inverse algebraic-geometric
spectral problem on one energy level for 2D Schr\"odinger
operator was posed and solved at the time when no direct spectral
theory was known. This theory was developed much later in \cite{Ksp},
where it was shown that the Bloch solutions for (\ref{3.11}) with
analytical periodic potential are parameterized by points of infinite
genus Riemann surface. It was proved that if this surface has finite genus,
then the Bloch functions have all the analytical properties suggested
in the inverse problem and therefore are just the Baker-Akhiezer-functions.
Moreover, it was proved that algebraic-geometric (finite-gap) potentials
are dense in the space of all periodic potentials.

\section{Spectral theory of operators with elliptic coefficients}

In this section we are going to discuss a specific spectral problem
for operators with elliptic coefficients, i.e. with the coefficients
that are meromorphic functions of a variable $x$ and have two
periods $2\omega, 2 \omega', {\rm Im} \ (\omega'/\omega)>0.$

Since Hermit time, it has been known that one-dimensional Schr\"odinger
operator with potential of the form $n(n+1)\wp(x)$  (where
$\wp(x)=\wp(x|\omega,\omega')$ is  Weierstrass $\wp$-function corresponding
to elliptic curve with periods $2\omega, 2 \omega'$, and $n$ is an integer)
has only $n$ gaps in the spectrum. These Lame potentials
had been the only known examples with the finite-gap property
before the finite-gap theory was constructed in the framework of the soliton
theory (see above). As we already have shown a generic algebraic-geometric
potential can be expressed in terms of higher genus Riemann $\theta$-function.
Sometimes higher genus formula can be reduced to exact expression in
terms of elliptic function.

The first example of this type which is different from
the Lame potentials was proposed in \cite{DN}.
Later a theory of elliptic finite-gap potentials attracted the special
interest due to remarkable observation made in \cite{amm} on their
connection with the elliptic Calogero-Moser model.
The most recent burst of interest is due to the unexpected
connections of these systems to Seiberg-Witten solution of $N=2$
supersymmetric gauge theories \cite{sw1,sw2}. It turns out that the low energy
effective theory for $SU(N)$ model with matter in the adjoint representation
(identified first in \cite{DW} with $SU(N)$ Hitchin system) is isomorphic to
the elliptic CM system. Using this connection quantum order parameters were
found in \cite{pd}.

The elliptic Calogero-Moser
(CM) system (\cite{c,m}) is a
system of $N$ identical particles on a line interacting with each other via
the potential $V(x)=\wp(x)$. Its equations of motion have the form
\beq
\ddot x_i= 4\sum_{j\neq i}\wp'(x_i-x_j). \label{cm}
\eeq
The CM system is a completely integrable Hamiltonian system, i.e. it
has $N$ independent integrals $H_k$ in involution (\cite{op}, \cite{op1}).
The second integral $H_2$ is the Hamiltonian of (\ref{cm}).

In \cite{amm} it was shown that the elliptic solutions of the
KdV equations have the form
\beq
u(x,t)=2\sum_{i=1}^N\wp(x-x_i(t))
\eeq
and the poles $x_i(t)$ of the solutions satisfy the constraint
$\sum_{j\neq i}\wp'(x_i-x_j)=0$,
which is the {\it locus} of the stationary points of the CM system. Moreover,
it turns out that the dependence of the poles with respect to $t$ coincides
with the Hamiltonian flow corresponding to the third integral $H_3$ of the
system. In \cite{kr3,chood} it was found that this connection becomes
an isomorphism in the case of the elliptic solutions of the
Kadomtsev-Petviashvilii equation. Moreover, in \cite{kr1} it was revealed
that the connection of the CM systems with the KP equation is in some
sense secondary and is a corollary of more fundamental connections
with spectral theory of {\it linear} operators with elliptic
potentials. The corresponding approach has been developed extensively
in \cite{bab,zab,wig}.

Let ${\cal L}$ be a linear differential or difference operator in
two variables $x,t$ with coefficients which are scalar or matrix elliptic
functions of the variable $x$. We do not assume any special
dependence of the coefficients with respect to the second variable.
Then it is natural to introduce a notion of {\it double-Bloch} solutions
of the equation
\beq
{\cal L} \psi=0. \label{gen}
\eeq
We call a {\it meromorphic} vector-function $f(x)$, which
satisfies the following monodromy properties:
\beq
f(x+2\omega_{\alpha})=B_{\alpha} f(x), \ \  \alpha=1,2,\label{g1}
\eeq
a {\it double-Bloch function}.  The complex numbers $B_{\alpha}$ are  called
{\it Bloch multipliers}.  (In other words, $f$ is a meromorphic section of a
vector bundle over the elliptic curve.)

In the most general form a problem that we are going to address is to
{\it classify} and to {\it construct} all the operators ${\cal L}$
such that equation (\ref{gen}) has {\it sufficiently enough} double-Bloch
solutions.

It turns out that existence of the double-Bloch solutions is so
restrictive that only in exceptional cases such solutions do exist.
A simple and general explanation of that is due to the Riemann-Roch
theorem. Let $D$ be a set of points $x_i,\ i=1,\ldots,m,$ on the elliptic
curve $\G_0$ with multiplicities $d_i$ and let $V=V(D; B_1,B_2)$ be a linear
space of the double-Bloch functions with the Bloch multipliers
$B_\a$ that have poles at $x_i$ of order less or equal to $d_i$ and
holomorphic outside $D$.
Then the dimension of $D$ is equal to:
$$
{\rm dim} \ D={\rm deg} \ D=\sum_i d_i.
$$
Now let $x_i$ depend on the variable $t$. Then for $f\in D(t)$ the function
${\cal L} f$ is a double-Bloch function with the same Bloch multipliers but in
general with higher orders of poles because taking derivatives and
multiplication by the elliptic coefficients increase these orders.
Therefore, the operator ${\cal L}$ defines a linear operator
$$
{\cal L}|_D: V(D(t);B_1,B_2)\longmapsto V(D'(t);B_1,B_2), \ N'=\deg D'>N=\deg D,
$$
and (\ref{gen}) is {\it always} equivalent to an {\it over-determined}
linear system of $N'$ equations for $N$ unknown variables which are the
coefficients $c_i=c_i(t)$ of expansion of $\Psi\in V(t)$ with respect to
a basis of functions $f_i(t)\in V(t)$. With some exaggeration one may say
that in the soliton theory the representation of a system  in the
form of the compatibility condition of an over-determined system of the
linear problems is considered equivalent to integrability.

In all of the basic examples $N'=2N$ and the over-determined system of
equations has the form
\beq
LC=kC,\ \ \p_tC=MC,\label{lax}
\eeq
where $L$ and $M$ are $N\times N$ matrix functions depending on a point $z$
of the elliptic curve as on a parameter. A compatibility
condition of (\ref{lax}) has the standard Lax form $\p_t L=[M,L]$, and is
equivalent to a finite-dimensional integrable system.

The basis in the space of the double-Bloch functions can be written in terms
of the fundamental function $\Phi(x,z)$ defined by the formula
\beq
\Phi(x,z)={\sigma(z-x)\over \sigma(z) \sigma(x)}
e^{\zeta(z)x}. \label{phi}
\eeq
Note, that $\Phi(x,z)$ is a solution of the Lame equation:
\beq
\left({d^2\over dx^2}-2\wp(x)\right)\Phi(x,z)=\wp(z)\Phi(x,z). \label{lame}
\eeq
>From the monodromy properties it follows that $\Phi$ considered
as a function of $z$ is double-periodic:
$$
\Phi(x,z+2\omega_{\alpha})=\Phi(x,z) ,
$$
though it is not elliptic in the classical sense due to an
essential singularity at $z=0$ for $x\neq 0$.

As a function of $x$ the function $\Phi(x,z)$ is double-Bloch function, i.e.
$$
\Phi(x+2\omega_{\alpha}, z)=T_{\alpha}(z) \Phi (x, z), \ T_{\alpha}(z)=
\exp \left(2\omega_{\a}\zeta(z)-2\zeta (\omega _{\alpha})z\right).
$$
In the fundamental domain of the lattice defined by
$2\omega_{\alpha}$ the function $\Phi(x,z)$ has a unique pole at the point
$x=0$:
\beq
\Phi(x,z)=x^{-1}+O(x). \label{j}
\eeq
The gauge transformation
$
f(x)\longmapsto \tilde f(x)=f(x)e^{ax},
$
where $a$ is an arbitrary constant does not change poles of any function and
transforms a double Bloch-function into another double-Bloch function. If $B_{\alpha}$
are Bloch multipliers for $f$ then  Bloch multipliers for $\tilde f$ are
equal to
\beq
\tilde B_1=B_1e^{2a\omega_1},\ \ \tilde B_2=B_2 e^{2a\omega_2}. \label{gn2}
\eeq
The two pairs of Bloch multipliers that are connected with each other
through the relation (\ref{gn2}) for some $a$ are called equivalent.
Note that for all equivalent pairs of Bloch multipliers the product
$
B_1^{\omega_2} B_2^{-\omega_1}
$
is a constant depending on the equivalence class, only.

>From (\ref{j}) it follows that a double-Bloch function $f(x)$ with simple
poles $x_i$ in the fundamental domain and with Bloch multipliers
$B_{\alpha}$ (such that at least one of them is not equal to $1$) may be
represented in the form:
\beq
f(x)=\sum_{i=1}^N c_i\Phi(x-x_i,z) e^{k x},\label{g2}
\eeq
where $c_i$ is a residue of $f$ at $x_i$ and $z$, $k$ are parameters
related by
$B_{\alpha}=T_{\alpha}(z) e^{2\omega_{\alpha}k }.$
(Any pair of Bloch multipliers may be represented in this form
with an appropriate choice of the parameters $z$ and $k$.)

Let us consider as an example the equation
\beq
{\cal L}\psi=(\p_t-\p_x^2+u(x,t))\psi=0 ,\label{cmo}
\eeq
where $u(x,t)$ is an elliptic function. Then as shown in \cite{kr1} equation
(\ref{cmo}) has $N$ linear independent double-Bloch solutions with equivalent
Bloch multipliers and $N$ simple poles at points $x_i(t)$ if
and only if $u(x,t)$ has the form
\beq
u(x,t)=2\sum_{i=1}^N\wp(x-x_i(t))     \label{u}
\eeq
and $x_i(t)$ satisfy the equations of motion of the elliptic CM system
(\ref{cm}).

The assumption that there exist $N$ linear independent double-Bloch
solutions with equivalent Bloch multipliers implies that they can be
written in the form
\beq
\psi=\sum_{i=1}^N c_i(t,k,z)\Phi(x-x_i(t),z)e^{kx+k^2t}, \label{psi}
\eeq
with the same $z$ but different values of the parameter $k$.

Let us substitute (\ref{psi}) into (\ref{cmo}). Then
(\ref{cmo}) is satisfied if and only if we get
a function holomorphic in the fundamental domain. First of all, we
conclude that $u$ has poles at $x_i$, only.
The vanishing of the triple poles $(x-x_i)^{-3}$ implies
that $u(x,t)$ has the form (\ref{u}). The vanishing of the double
poles $(x-x_i)^{-2}$ gives the equalities that can be written as a matrix
equation for the vector $C=(c_i)$:
\beq (L(t,z)-kI)C=0\,,
\label{l}
\eeq
where $I$ is the unit matrix and the Lax matrix $L(t,z)$ is defined as
follows:
\beq
L_{ij}(t,z)=-{1\over 2}\delta_{ij}\dot x_i-(1-\delta_{ij})\Phi(x_i-x_j,z).
\label{L}
\eeq
Finally, the vanishing of the simple poles gives the equations
\beq (\partial_t-M(t,z))C=0\,,  \label{m} \eeq where \beq
M_{ij}=\left(\wp(z)-2\sum_{j\neq
i}\wp(x_i-x_j)\right)\delta_{ij}-2(1-\delta_{ij})\Phi' (x_i-x_j ,z). \label{M}
\eeq
The existence of $N$ linear independent
solutions for (\ref{cmo}) with equivalent Bloch multipliers implies that
(\ref{l}) and (\ref{m}) have $N$ independent solutions corresponding to
different values of $k$. Hence, as a compatibility condition we get the Lax
equation $\dot L=[M,L]$ which is equivalent to (\ref{cm}). Note that the last
system does not depend on $z$. Therefore, if (\ref{l}) and (\ref{m}) are
compatible for some $z$, then they are compatible for all $z$. As a result we
conclude that if (\ref{cmo}) has $N$ linear independent double-Bloch solutions
with equivalent Bloch multipliers then it has infinitely many of them. All the
double-Bloch solutions are parameterized by points of an algebraic curve $\G$
defined by the characteristic equation
\beq
R(k,z)\equiv \det (kI-L(z))=k^N+\sum_{i=1}^Nr_i(z)k^{N-i}=0. \label{r} \eeq
Equation (\ref{r}) can be seen as a dispersion relation between two Bloch
multipliers and defines $\G$ as $N$-sheet cover of $\G_0$.

As was shown in \cite{kr1} expansion of the characteristic equation (\ref{l})
at $z=0$ has the form:
\beq
R(k,z)=\prod_{i=1}^N(k+\nu_iz^{-1}+h_i+O(z)),\ \ \nu_1=1-N, \ \nu_i=1, \ i>1.
\label{r1}
\eeq
We call the sheet of $\G$ at $z=0$ corresponding to the branch
$k=z^{-1}(N-1)+O(1)$, an upper sheet and mark the point $P_1$ on this
sheet among the preimages of $z=0$. From (\ref{r1}) it follows that in general
position when the curve $\G$ is smooth, its genus equals $N$.

Further consideration of analytical properties of the function $\psi$ given
by (\ref{psi}) where $c_i$ are components of the eigenvector to the matrix $L$
shows that this function is just the Baker-Akhiezer function introduced in
Section 6. Combined with expression (\ref{f1}) for $\psi$ in terms of
$\theta$-function this result directly
leads to the main statement of \cite{kr1}:

{\it the coordinates of the particles $x_i(t)$ are roots of the equation}
\beq
\theta(Ux+Vt+Z_0)=0, \label{for}
\eeq
where $\theta(\xi)=\theta(\xi|B)$ is the Riemann theta-function
corresponding to the  matrix of $b$-periods of holomorphic differentials on $\G$;
the vectors $U$ and $V$ are the vectors of $b$-periods of normalized
meromorphic differentials on $\G$, with poles of order 2 and 3 at the point
$P_1$.

Among other examples of integrable systems that can be generated in the
similar way are Ruijesenaars-Schneider system \cite{rs}
\beq
\ddot x_i = \sum_{s\neq i} \dot
x_i \dot x_s (V(x_i-x_s)-V(x_s-x_i)), \
V(x)=\zeta(x)-\zeta(x+\eta),
\label{rs}
\eeq
and nested Bethe ansatz equations  \cite{nj}
\beq
\prod_{j\neq i}{\s(x_i^n-x_j^{n+1})\s(x_i^n-\eta-x_j^n)\s(x_i^n-x_j^{n-1}+\eta)
\over \s(x_i^n-x_j^{n-1})\s(x_i^n+\eta-x_j^n)\s(x_i^n-x_j^{n+1}-\eta)}=-1
\label{bet}
\eeq
As shown in \cite{zab,wig}, they are generated by spectral problems for
equations
\beq
{\cal L}\psi=\p_t\psi(x,t)-\psi(x+\eta,t)-v(x,t)\psi(x,t)=0, \label{41}
\eeq
and
\beq
\psi(x,m+1)=\psi(x+\eta)+v(x,m)\psi(x,m) \label{51},
\eeq
respectively.  (Here $\eta$ is a complex number and $v(x,t)$ is an elliptic
function.)

Strange as it is, the inverse spectral problem which is discussed here
is simplier for two-dimensional operators then for one-dimensional stationary
operators. For example a family of spectral curves corresponding
to operators (\ref{cmo}) that have double-Bloch solutions can be
described explicitly. A nice formula was found in \cite{pd}:
\beq
R(k,z)=f(k-\zeta(z),z), \ \ f(k,z)={1\over \s(z)}\ \s
\left(z+{\p\over \p k}\right)H(k),
\label{ph}
\eeq
where H(k) is a polynomial.
Note that (\ref{ph}) may be written as:
$$
f(k,z)={1\over \s(z)}\ \sum_{n=1}^N {1\over n!}\p_z^n \s(z)
\left({\p\over \p k}\right)^n H(k).
$$
The coefficients of the polynomial $H(k)$ are free parameters of the spectral
curve of the CM system.

The spectral curves corresponding to the Schr\"odinger operator
with the same property are special case of the curves (\ref{ph}) but
their explicit description is unknown. In particular, exact formula
for branching points for Lame potentials is unknown. As it was mentioned
above the first example of elliptic finite-gap potentials different
from the Lame potentials was found in \cite{DN}. A wide class of such
potentials was found in \cite{VT}.

In \cite{ges} it was noted that the problem of classification
of Schr\"odinger operators with elliptic potentials that have
two double-Bloch solutions for almost all energy levels was posed
by Picard though had not been solved until very recently.
In \cite{ges} using Floquet spectral theory for
the Schr\"odinger operator it was proved that all such potentials are
finite-gap. This result is an essential step in the Picard problem
though its complete and effective solution is still an open problem.

\end{document}